\documentclass[a4paper,fleqn,usenatbib]{mnras}

\usepackage{mathptmx}

\usepackage[T1]{fontenc}
\usepackage{ae,aecompl}

\usepackage{graphicx}	% Including Figure files
\usepackage{amsmath}	% Advanced maths commands
\usepackage{amssymb}	% Extra maths symbols
\usepackage{pdflscape}
\usepackage{rotating}

\newcommand{\kms}{km s$^{-1}$}

\title[Mapping Observations of COMs around Sgr B2]{Mapping Observations of complex organic molecules around Sagittarius B2 with ARO 12m telescope}

\author[Juan Li et al.]{Juan Li,$^{1,2}$\thanks{E-mail: lijuan@shao.ac.cn}
  Junzhi Wang,$^{1,2}$
  Haihua Qiao,$^{3,1,2}$
  Donghui Quan,$^{4,5}$
  Min Fang,$^{6}$  
  \newauthor
  Fujun Du,$^{2,7}$
  Fei Li,$^{1,2,8}$
  Zhiqiang Shen,$^{1,2}$
  Shanghuo Li,$^{1,2,8}$
  Di Li,$^{9,10}$   
  \newauthor
  Yong Shi,$^{11,12,13}$
  Zhiyu Zhang,$^{11,12,13}$
  and
  Jiangshui Zhang,$^{14}$
\\
\\
$^{1}$Department of Radio Science and Technology, Shanghai Astronomical observatory, 80 Nandan RD, Shanghai 200030, China \\
$^{2}$Key Laboratory of Radio Astronomy, Chinese Academy of Sciences, China  \\
$^{3}$National Time Service Center, Chinese Academy of Sciences, Xi'An, Shaanxi, 710600, PR China  \\
$^{4}$Xinjiang Astronomical Observatory, Chinese Academy of Sciences, 150 Science 1-Street, Urumqi 830011, PR China  \\
$^{5}$Department of Chemistry, Eastern Kentucky University, Richmond, KY 40475, USA  \\
$^{6}$Department of Astronomy, University of Arizona, 933 North Cherry Avenue, Tucson, AZ 85721, USA  \\
$^{7}$Purple Mountain Observatory, Chinese Academy of Sciences, Nanjing 210034, PR China  \\
$^{8}$University of Chinese Academy of Sciences, 19A Yuquanlu, Beijing 100049, PR China  \\
$^{9}$ National Astronomical Observatories, Chinese Academy of Sciences, A20 Datun Road, Chaoyang District, Beijing 100012, PR China  \\
$^{10}$ CAS Key Laboratory of FAST, Beijing, CN
$^{11}$School of Astronomy and Space Science, Nanjing University, Nanjing 210093, China  \\
$^{12}$Key Laboratory of Modern Astronomy and Astrophysics (Nanjing University), Ministry of Education, Nanjing 210093, China  \\
$^{13}$Collaborative Innovation Center of Modern Astronomy and Space Exploration, Nanjing 210093, China    \\
$^{14}$Center for Astrophysics, Guangzhou University, Guangzhou, 510006, PR China    
}

\date{Accepted XXX. Received YYY; in original form ZZZ}

\pubyear{2019}

\begin{document}
\label{firstpage}
\pagerange{\pageref{firstpage}--\pageref{lastpage}}
\maketitle

\begin{abstract}

We performed high-sensitivity mapping observations of several complex organic molecules around Sagittarius B2 with ARO 12m telescope at 3-mm wavelength. Based on their spatial distribution, molecules can be classified as either ``extended" that detected not only in Sgr B2(N) and Sgr B2(M), or ``compact" that only detected toward or near to Sgr B2(N) and Sgr B2(M). The ``extended" molecules including glycolaldehyde (CH$_2$OHCHO), methyl formate (CH$_3$OCHO), formic acid (t-HCOOH), ethanol (C$_2$H$_5$OH) and methyl amine (CH$_3$NH$_2$), while the ``compact" molecules including dimethyl ether (CH$_3$OCH$_3$), ethyl cyanide (C$_2$H$_5$CN), and amino acetonitrile (H$_2$NCH$_2$CN). These ``compact" molecules are likely produced under strong UV radiation, while ``extended" molecules are likely formed under low-temperature, via gas-phase or grain surface reactions. The spatial distribution of ``warm" CH$_2$OHCHO at 89 GHz differ from the spatial distribution of ``cold" CH$_2$OHCHO observed at 13 GHz. We found evidence for an overabundance of CH$_2$OHCHO compared to that expected from the gas-phase model, which indicates that grain-surface reactions are necessary to explain the origin of CH$_2$OHCHO in Sagittarius B2. Grain-surface reactions are also needed to explain the correlation between the abundances of ``cold" CH$_2$OHCHO and C$_2$H$_5$OH. These results demonstrate the importance of grain-surface chemistry in the production of complex organic molecules. 

\end{abstract}


\section{Introduction}

Complex organic molecules (COMs) are regarded to be related to the origin of life on Earth \citep{herbst2009}. Thus, understanding property and formation mechanism of prebiotic molecules is key for study of astrobiology.  It has been found that prebiotic molecules are very abundant in the Galactic center giant molecular cloud Sagittarius B2 (Sgr B2) \citep{menten2011}. Sgr B2 contains two main sites of star formation, Sgr B2(N) and Sgr B2(M), which have been the best hunting ground for prebiotic molecules in the interstellar medium (ISM) since the early 1970s \citep{belloche2013}. Many COMs are first detected toward Sgr B2(N), such as glycoaldehyde CH$_2$OHCHO \citep{hollis2000}, ethylene glycol HOCH$_2$CH$_2$OH \citep{hollis2002}, amino acetonitrile H$_2$NCH$_2$CN \citep{belloche2008} and so on. Some COMs, such as glycolaldehyde and ethylene glycol, were found to be widespread in the Galactic center \citep{hollis2001, requena2006, requena2008, li2017}.

Theoretical and experimental studies \citep{garrod2008, woods2013, skouteris2018, oberg2009, fedoseev2015, chuang2016, meinert2016} show that COMs may be formed on cosmic dusts by radicals that are produced on dust grain surfaces by UV or cosmic ray induced photons. The COMs are then desorbed into the gas phase due to thermal and/or non-thermal processes \citep{garrod2006}. However, these complex ice compositions cannot be directly observed with telescopes in the infrared; thus, most constraints come from millimeter observations of desorbed ice chemistry products. On the other hand, recent quantum chemical computations show that the combination of radicals trapped in amorphous water ice does not necessarily lead to the formation of COMs \citep{enrique2016}. Recently, a new scheme for the gas-phase synthesis of glycolaldehyde, a species with a prebiotic potential, has been proposed. The predicted abundance of glycolaldehyde agrees well with that measured in solar-type hot corinos and shock sites \citep{skouteris2018}. Thus, how these species are synthesized is a mystery.

%A comparison of the spatial distribution of different COMs would shed light on the chemical models. 
Previous mapping observations mainly concentrate on Sgr B2(N) with interferometers like ATCA and ALMA at small spatial scales ($\leq$ 1 \arcmin) \citep{corby2015, sanchez2017, belloche2016, belloche2019, bonfand2019, xue2019}, and only a handful of large-scale spectral line mapping observations toward Sgr B2 has been performed \citep{jones2008, jones2011, chengalur2003}. The large-scale spatial distribution of COMs like C$_2$H$_5$CN, HCOOH are still unclear mainly because of their weak emission \citep{halfen2017}. Single dish telescopes have ability to probe the structure of weak lines from extended gas of Sgr B2 complex on spatial scales much larger than 1\arcmin. In this paper we present high-sensitivity grid imaging results of several COMs around Sgr B2 with the Arizona Radio Observatory (ARO) 12 m telescope. In \S2, we describe the observations and data reduction. In \S3, we present the mapping result, and spatial distribution of column densities, rotational temperatures, and estimated abundances for the observed COMs. In \S4, we discuss molecular formation pathways and compare our results with protostars and then summarize our conclusions in \S5.

\section{Observations and data reduction}

We performed point by point mapping observations of COMs at 3-mm wavelength around Sgr B2 in 2017 April and May with the Arizona Radio Observatory 12 m telescope at Kitt Peak, Arizona. The 85 - 116 GHz Sideband Separating ALMA Band-3 Receiver was used. The beam size is 70 arcsec at 90 GHz. The 2 IF modes of Millemeter Auto Correlator (MAC) were employed as backends. The MAC has 4096 channels with 195 kHz sampling interval, corresponding to 0.65 \kms at 90 GHz. Though the velocity resolution is not high enough for study of molecular emission from hot cores, but enough for single dish observations of Sgr B2. The FWHM linewidth is up to $\sim$ 20 km s$^{-1}$ \citep{rivilla2018, rivilla2019} due to the extended envelope and several unresolved cores with different velocities and linewidths within the beam. The temperature scale was determined by the chopper wheel method, and is related to the main beam brightness temperature with main beam efficiency of 0.88. The system temperature range from 90 to 150 K. The integration time for each position range from 50 minutes to 90 minutes. 

Observations were conducted in position-switching mode, and an offset of +30 arcmin in azimuth was used. The data processing was conducted using \textbf{GILDAS} software package\footnote{\tt http://www.iram.fr/IRAMFR/GILDAS.}, including CLASS and GREG. Linear baseline subtractions were used for most of the spectra. For each transition, the spectra of subscans, including two polarizations, were averaged to reduce rms noise levels. Gaussian fitting is used to derive the physical properties of molecule lines, including peak intensity, V$_{LSR}$, FWHM line width, and integrated intensity.

\section{Observing Results}

\begin{figure*}
\centering
\includegraphics[width=2.6in, angle=90]{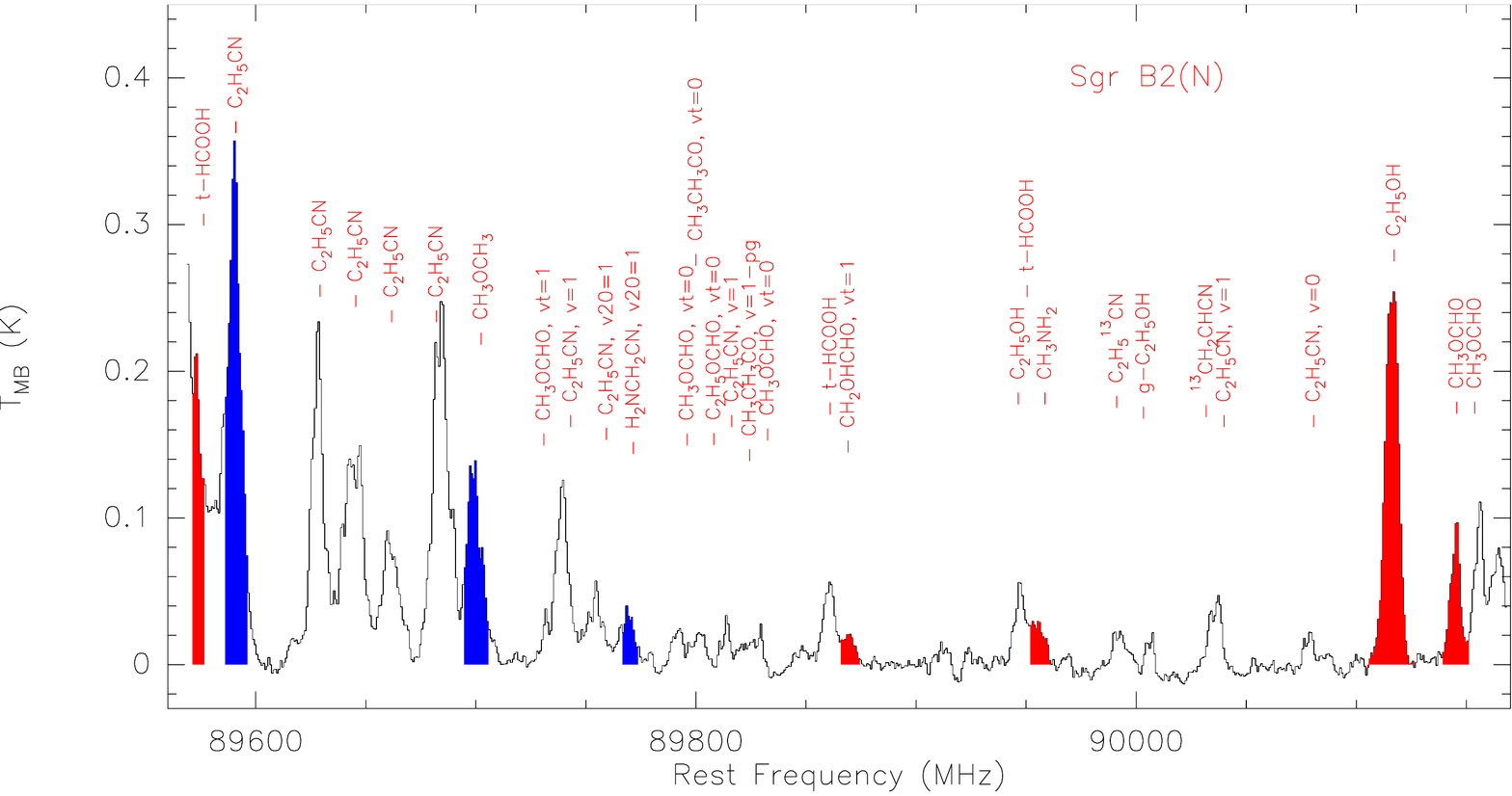}
\includegraphics[width=2.6in,angle=90]{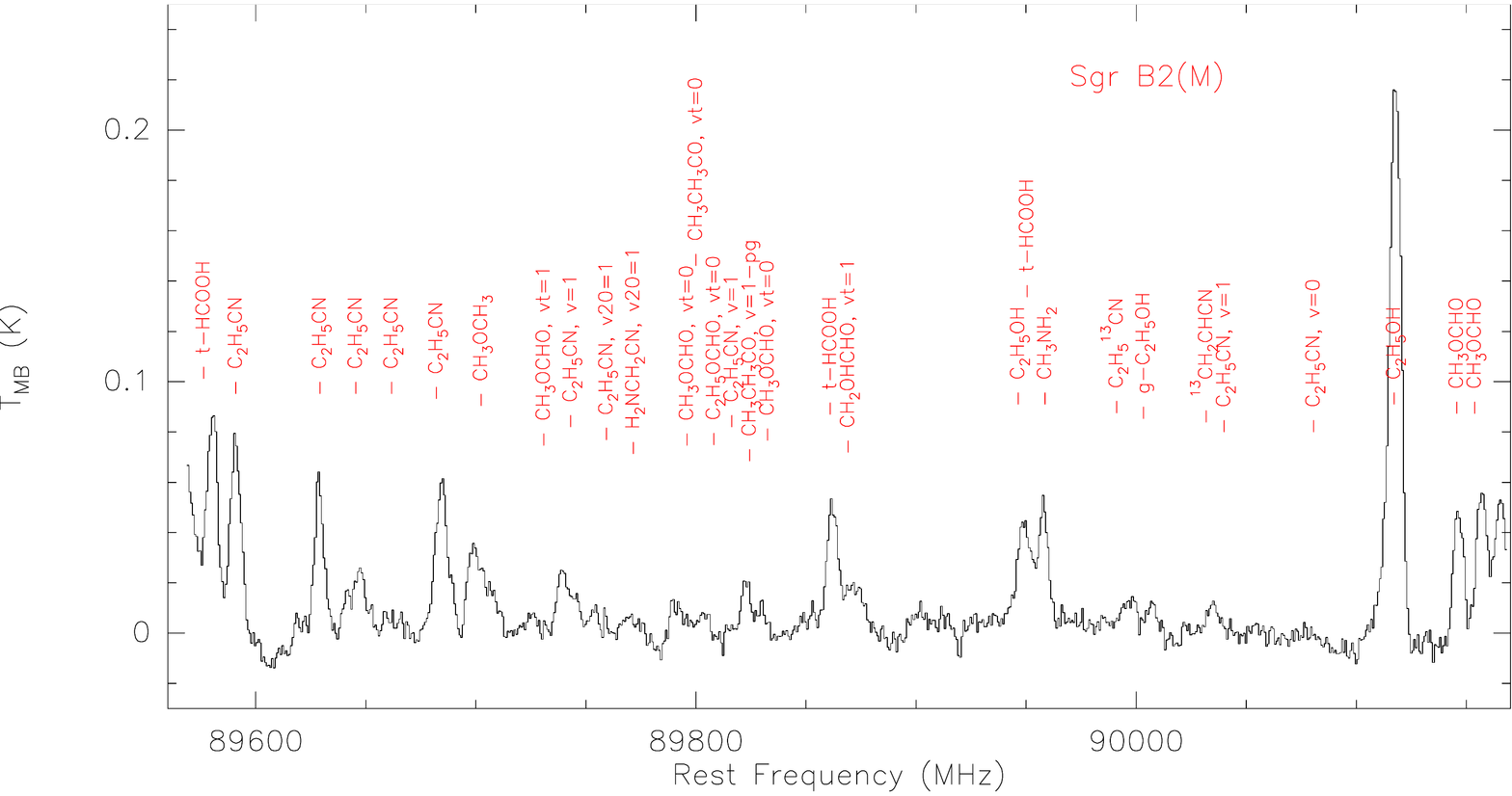}
\includegraphics[width=2.6in,angle=90]{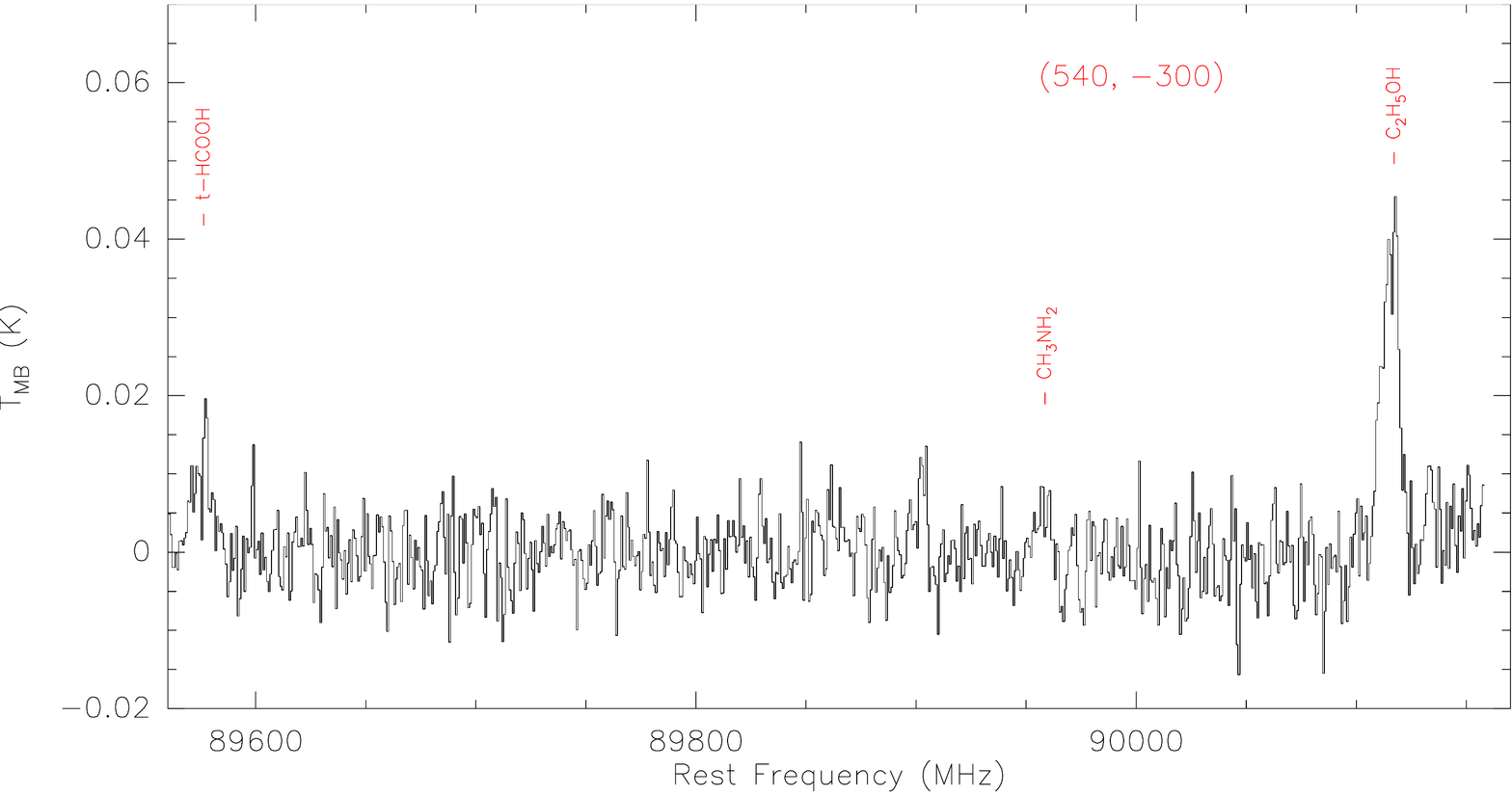}
\caption{Spectra obtained toward Sgr B2(N), Sgr B2(M) and (540, -300) away from Sgr B2(N) with the ARO 12m telescope in the main-beam temperature scale. The line names are labeled. The ``extended" molecular line profiles were colored with red, while the ``compact" molecular line profiles were colored with blue in spectra of Sgr B2N.}
\label{fig 1}
\end{figure*}

We detected emission lines from t-HCOOH, C$_2$H$_5$CN, CH$_3$OCH$_3$, CH$_3$OCHO, H$_2$NCH$_2$N, CH$_2$OHCHO, CH$_3$NH$_2$, and C$_2$H$_5$OH toward many positions at a rms level of $\sim$ 3 mK. Figure \ref{fig 1} presents spectra observed toward several positions with ARO 12m telescope. As is shown in Figure \ref{fig 1}, some molecule lines are only detected toward Sgr B2(N) and Sgr B2(M). Table \ref{table 1} presents detected transitions, and their spectroscopic properties in Sgr B2(N). The molecular spectroscopic parameters are taken from the current public databases: the CDMS catalog\footnote{\tt https://www.astro.uni-koeln.de/cdms/catalog} \citep{muller2005}, the JPL catalog\footnote{\tt https://spec.jpl.nasa.gov} \citep{pickett1998}
 and the Spectral Line Atlas of Interstellar Molecules (SLAIM) database, which are available in the SPLATALOGUE spectroscopy database\footnote{\tt https://www.splatalogue.net.}. Note that the strong C$_2$H$_5$CN emission near to t-HCOOH 4(0,4)-3(0,3) (89.57917 GHz) make it difficult to determine the baseline and intensity of t-HCOOH. Only t-HCOOH emission toward Sgr B2(N) was affected since the C$_2$H$_5$CN emission was very weak toward other positions.

\begin{table*}
\scriptsize
    \begin{center}
      \begin{minipage}{160mm}
      \caption{Identified Transitions and their observed parameters in Sgr B2(N)}
      \label{table 1}
      \begin{tabular}{lccccccccccccc}
    \hline
    \hline
%              &                  &                                 &                  &                     &                       &    Sgr B2(N)    &                  &                    &                            &                               \\
 Species & Transitions &    Rest Freq.   &   $E_{u}$  & $\mu^2$S    &     $T_{mb}$  & V$_{LSR}$    &   $\Delta V$   &  $\int T_{mb}$ dv      \\
             &          &   (GHz)                       &     (K)      & (D$^2$)  &  (mK)      & (km s$^{-1}$)   &     (km s$^{-1}$)   &    (mK km s$^{-1}$)     \\
\hline        
          t-HCOOH$^*$          &    4(0,4)-3(0,3)       &    89.57917(1e-6)           &    10.76      &     8.077     & $\sim$100   &   $\sim$64   &   $\sim$20    & $\sim$2000               \\      %    map
   C$_2$H$_5$CN,v=0 & 10(9,1)-9(9,0)  &     89.584987(5e-5)        &   109.32          &    28.16    & blended & - &   - &  -        \\      %                   
   C$_2$H$_5$CN,v=0 & 10(9,2)-9(9,1)  &     89.584987(5e-5)        &   109.32          &    28.16    & blended & - &   - &  -        \\      %                   
  C$_2$H$_5$CN,v=0$^*$ & 10(4,7)-9(4,6)   &     89.590028(3e-6)     &   41.44          &    124.52  & 160(45) & 63.0(.6) &   22.0(1.4) &  3520(610)      \\         %  peak:   0.3 K map
   C$_2$H$_5$CN,v=0 & 10(4,6)-9(4,5)  &     89.591013(3e-6)        &   41.44          &    124.52    & 160(45) & 63.0(.6) &   22.0(1.4) &  3520(610)        \\      %                   
       C$_2$H$_5$CN,v=0 & 10(3,8)-9(3,7)  &  89.628485(3e-6)      &   33.66              &    134.89     & 200(10)  & 65.0(1.4) &  21.9(4.0) & 4666(654)       \\      %    0.2 K
     C$_2$H$_5$CN,v20=1 & 10(8,3)-9(8,2) &  89.643412(2e-6)               &  630.76     &   52.43     &  blended   &          &           &                 \\      %    0.1 K                              
     C$_2$H$_5$CN,v20=1  & 10(4,7)-9(4,6)  &   89.661002(5e-5)     &    577.94    &   122.32    & 30(3) & 67.2(.4)  & 7.0(1.2) &  211(57)           \\      %    0.05 K                   
    C$_2$H$_5$CN,v=0   & 10(3,7)-9(3,6) &  89.684710(3e-6)     &    33.66    &  134.88    &  134(3)   &   60.9(2.6)  &  26.8(2.6)  &  3821(236)          \\            %       0.2 k                                    
   CH$_3$OCH$_3$,v=0  &  2(2,1)-2(1,2)AE &   89.697737(1e-6)     &      8.36      &    8.475      &  91(8) & 66.5(2.6)  &  18.4(2.6)  & 1783(158)          \\         %              
    CH$_3$OCH$_3$,v=0$^*$ & 2(2,1)-2(1,2)EE &  89.699809(1e-6)         &   8.36          &    22.31    & 50(8) & 64.2(2.6) & 15.1(2.6)  & 797(158)         \\      %       0.1 k      map
    CH$_3$OCH$_3$,v=0 &  2(2,1)-2(1,2)AA   &  89.702829(2e-6)         &   8.36       &   14.131   & 46(8) & 65.2(.3)  &  23.0(2.6) & 1131(158)            \\      %       0.1 k                     
    CH$_3$OCHO, vt=1 &   8(0, 8) - 7(0, 7) A   &  89.731695(1e-5)  &      207.83      &   41.664      & 29(6)  &  64.2(.3)  & 7.4(.7)  &  228(19)      \\          %   0.02 k
%  C$_2$H$_5$CN,v=1  & 10(4,7)-9(4,6) E   &  89.739675(3e-6)	  &   328.23       &   30.007      &  110(6)  & 66.1(.3) & 21.4(.7)  & 2507(66)        \\      %    0.1 k
%    C$_2$H$_5$CN,v=1  & 10(4,6)-9(4,5)0   &  89.738635(1e-6)	  &   332.93       &   31.083      &  blended  &   &      &          \\
 %  C$_2$H$_5$CN,v=1  & 10(4,6)-9(4,5)2   &  89.739013(3e-6)	  &   332.93       &   30.007      &  blended  &   &      &          \\
  %    C$_2$H$_5$CN,v=1  & 10(4,7)-9(4,6)1   &  89.739789(3e-6)	  &   332.93       &   30.007      &  blended  &   &      &          \\
   %      C$_2$H$_5$CN,v=1  & 10(4,7)-9(4,6)1   &  89.740344(1e-6)	  &   332.93       &   31.082      &  blended  &   &      &          \\
   C$_2$H$_5$CN,v20=1 &   10(3,7) - 9(3,6)   &  89.754402(5e-5)     &     570.24     &     132.51    &  43(5)  &  65.9(.4) &  34.3(1.2) & 1579(46)      \\      %     0.05 k
   H$_2$NCH$_2$CN$^*$  &  10(0,10)-9(3,6) &    89.770285(7e-6)     &     19.50      &    66.296     &  25(3)  &  63.7(0.7) & 14.5(1.6) &  377(40)        \\         %    0.02 k,blended with U line                                        
   t-HCOOH            &   4(2,3)-3(2,2)      &   89.861484(2e-6)     &       23.51    &     5.93      &  50(3)  & 66.9(.2)  & 18.5(.6)  & 987(26)        \\      %  0.05 k 
  CH$_2$OHCHO$^*$  &   9(1,8)-8(2,7)   &   89.868639(6e-6)    &       26.43   &  20.565 &  17(3)  &  61.5(2.6)  & 18.6(2.6)  &  338(41)      \\         %   0.02 k                 
 t-HCOOH            &   4(3,2)-3(3,1)       &    89.94821(2e-6)              &     35.09  &  3.536  & blended    &               &               &               \\           %    0.05 ? blended  !!!                                      
   CH$_3$NH$_2$$^*$ &    1(1)A1-1(0)A2       &   89.956068(2e-5)        &    6.45     &    1.5$^a$      &  31(3)  &  67.8(1.3)  & 30.3(2.8)  &  982(87)        \\      %     0.04 K, map
   CH$_3$CH$_2$C$^{13}$N  &   10(2,8)-9(2,7)         &     89.992750(1.2e-6)            &     28.10         &    139.763      &     blended           &               &               &                        \\      %   0.03 k 
    g-C$_2$H$_5$OH  &  13(1,12)-13(1,13),g+ &   90.0077(5e-5)     &  135.11     &  13.99     &  28(3)   &  64.0(.5)  &   6.2(1.2) & 188(53)            \\        %     0.03 k                                                                                                               
    $^{13}$CH$_2$CHCN &   10(1,10)-9(1,9)  &   90.032588(1.2e-6) &   25.92      &   432.232     &    blended  &               &               &                        \\       %    0.04 k                   
    CH$_3^{13}$CH$_2$CN, v=0  &  10(2,8)-9(2,7)   &       90.079288(1.3e-6)        &   28.00       &  139.77   &     24(3)     &  66.3(6) &  17.7(1.3) &  457(29)       \\         %      0.03 k                               
  C$_2$H$_5$OH$^*$  &  4(1,4)-3(0,3)       &    90.11761(2.8e-6)          &         9.36                &   5.21       & 266(9)  &  68.5(.1) &  22.4(.2)   & 6330(37)          \\      %    0.3 k           map
  CH$_3$OCHO$^*$   &   7(2,5)-6(2,4) E       &    90.145634(2.4e-5)       &        19.68           &   34.275  &  86(7)   &  65.7(.1)  &  16.7(.4)  & 1523(28)               \\            %       0.1 k     map
 CH$_3$OCHO   &   7(2,5)-6(2,4) A    &    90.156473(1e-5)       &        19.67       &   34.289   &  80(7)  &  66.3(.4) &  13.7(.9)  &  1166(65)               \\            %       0.1 k                                                  
\hline
      \end{tabular}  \\
    \end{minipage}
  \end{center}
  Notes.- $^*$: Lines used for mapping. Col. (1): chemical formula; Col. (2): transition quantum numbers, Col. (3): rest frequency; Col. (4): upper state energy level (K); Col. (5): dipole-weighted transition dipole matrix elements; Col. (6): peak temperature; Col. (7): centroid velocity; Col. (8): FWHM linewidth; Col. (9): integrated intensity.
       $^a$: $\mu^2$S is taken from \citep{motiyenko2014} for CH$_3$NH$_2$. 
\end{table*}

\begin{figure*}
\centering
\includegraphics[width=2.3in, angle=-90]{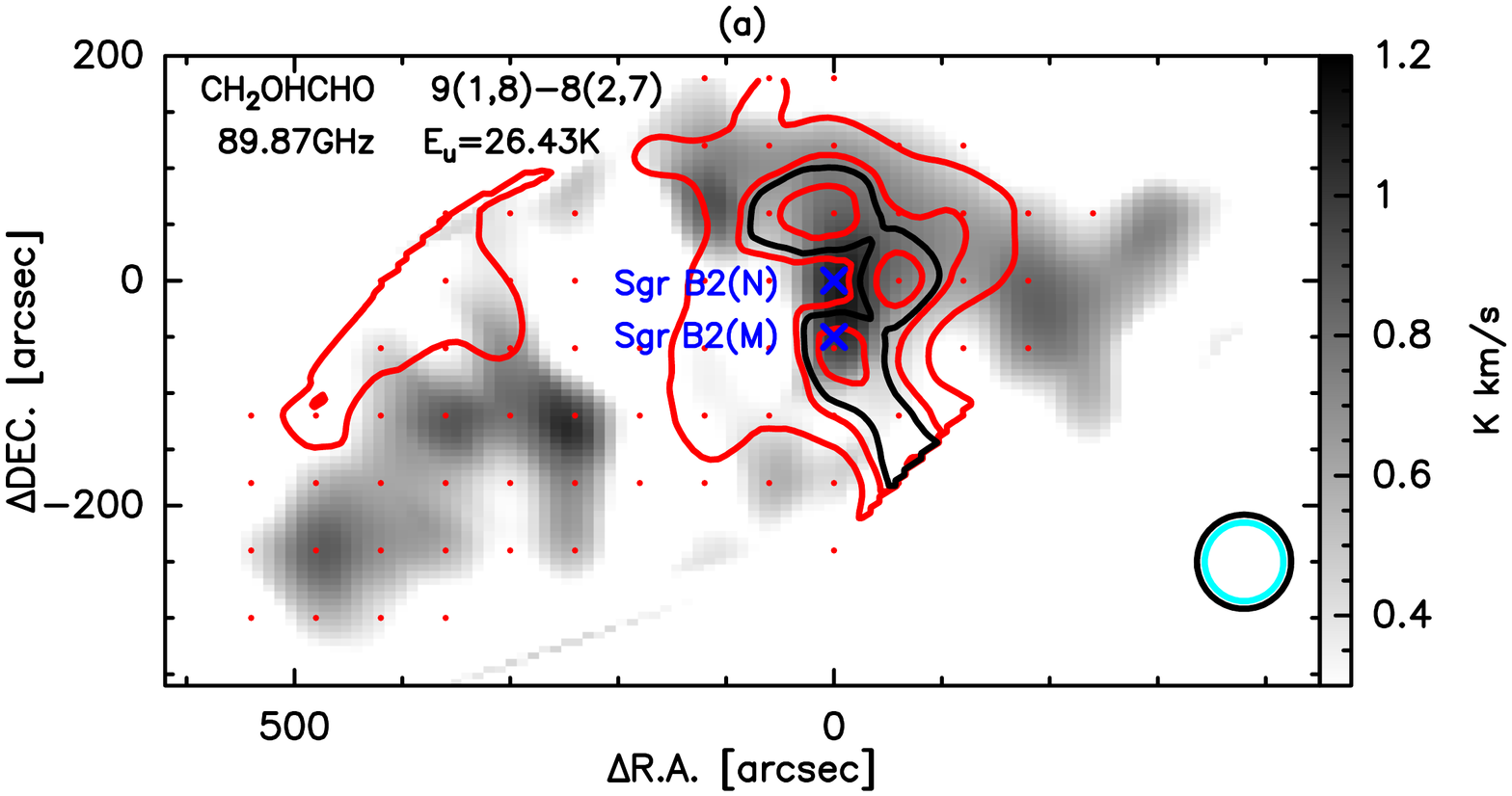}
\includegraphics[width=2.3in, angle=-90]{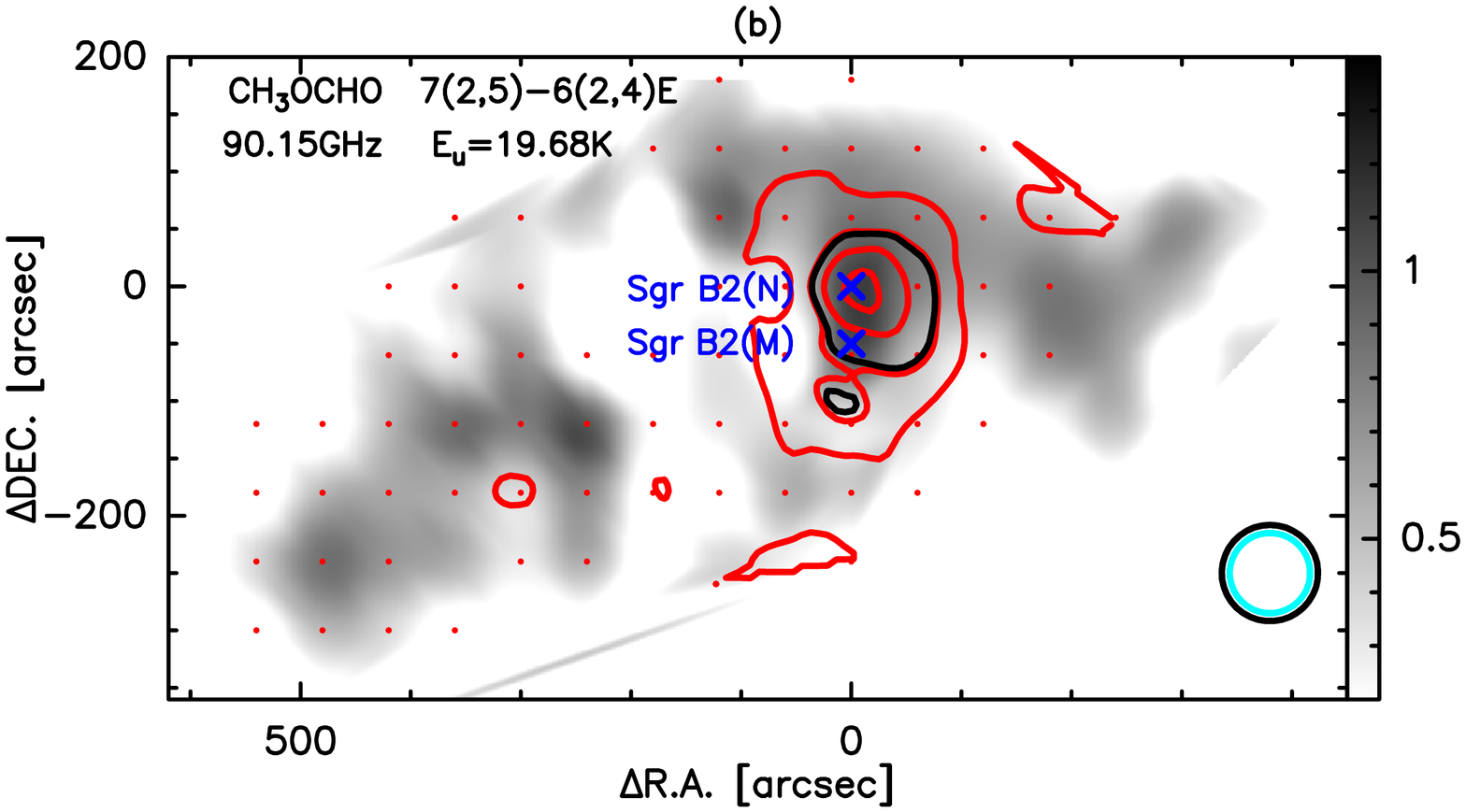}
\includegraphics[width=2.3in, angle=-90]{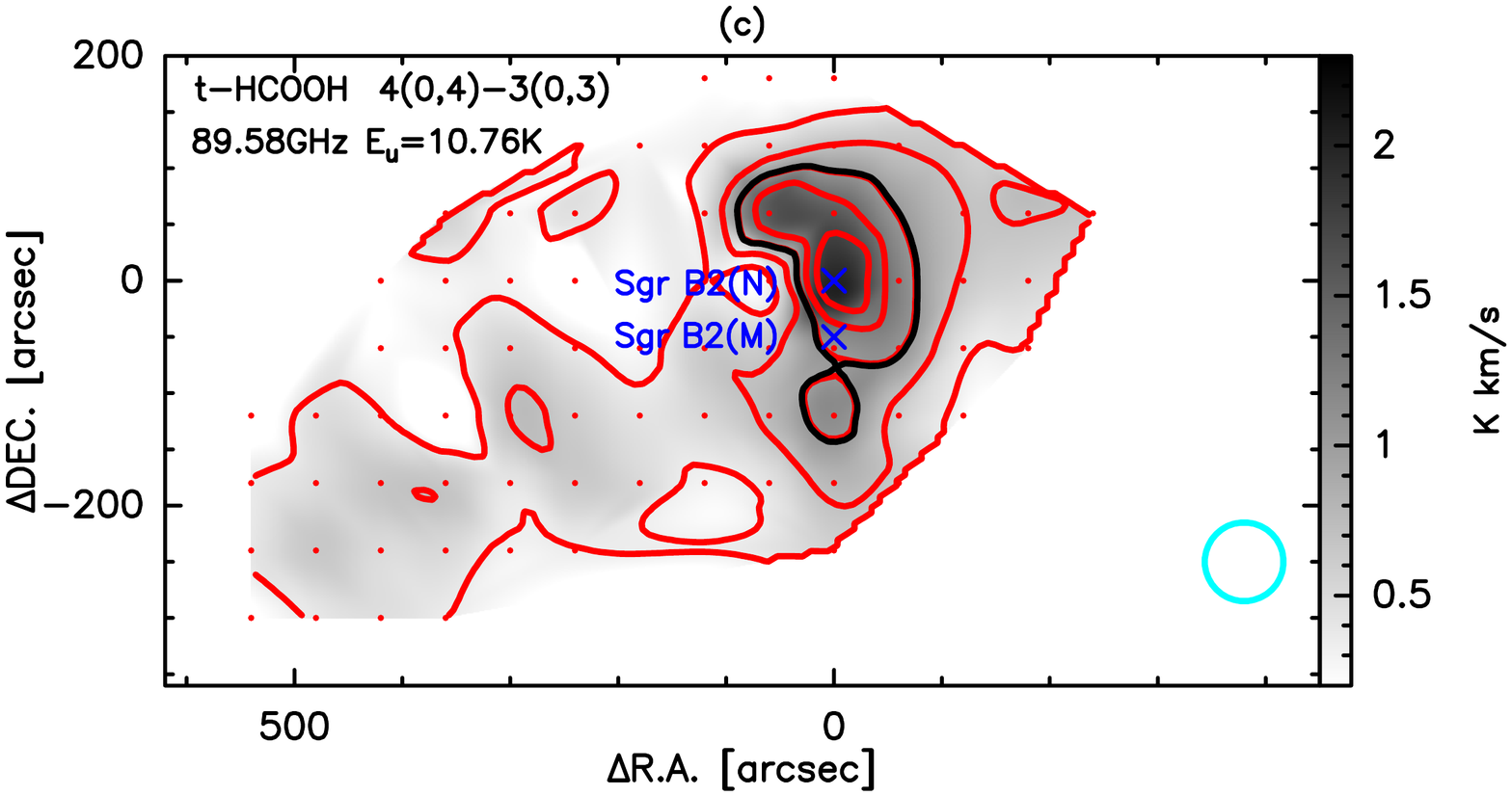}
\includegraphics[width=2.3in, angle=-90]{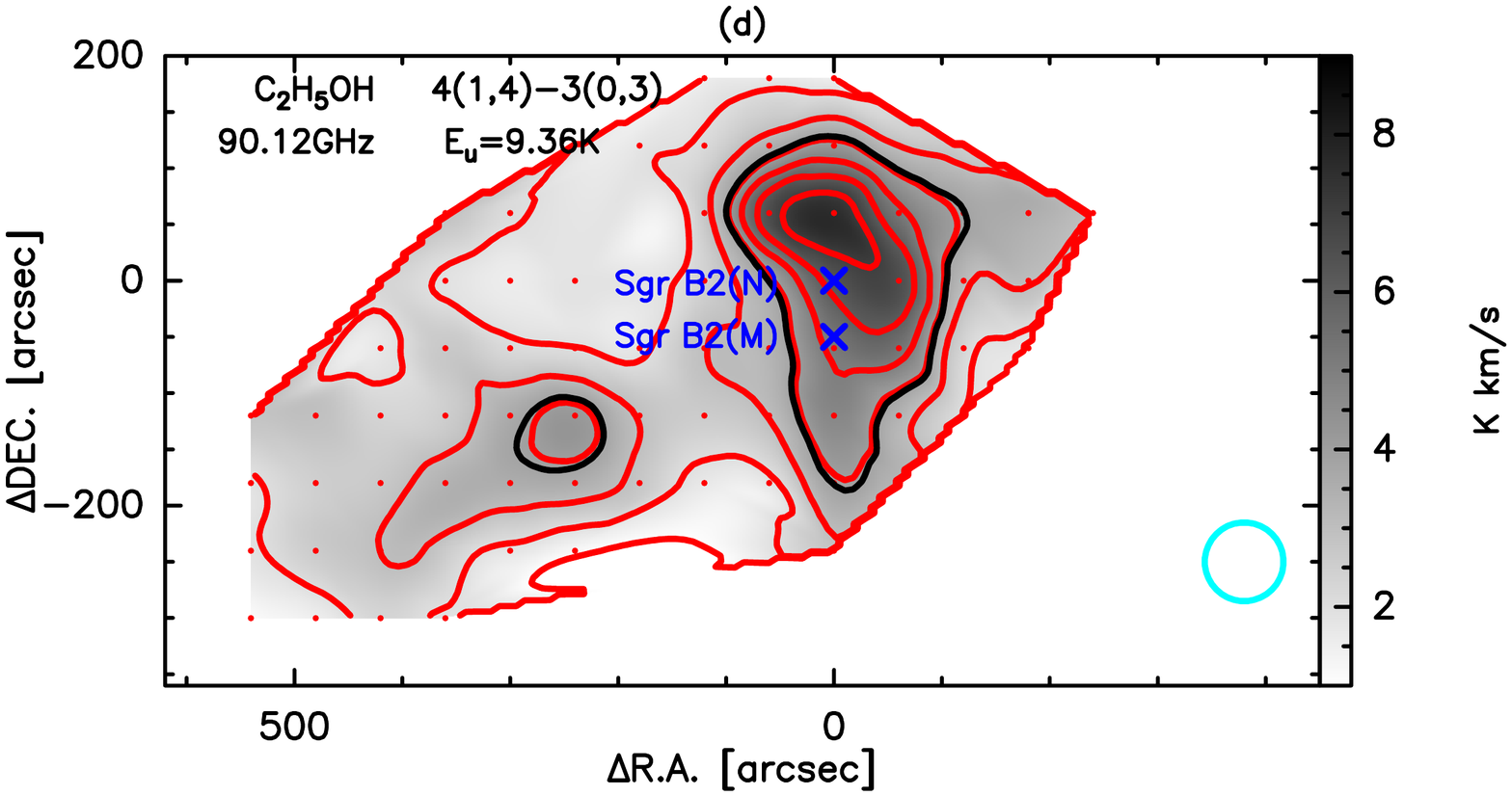}
\includegraphics[width=2.3in, angle=-90]{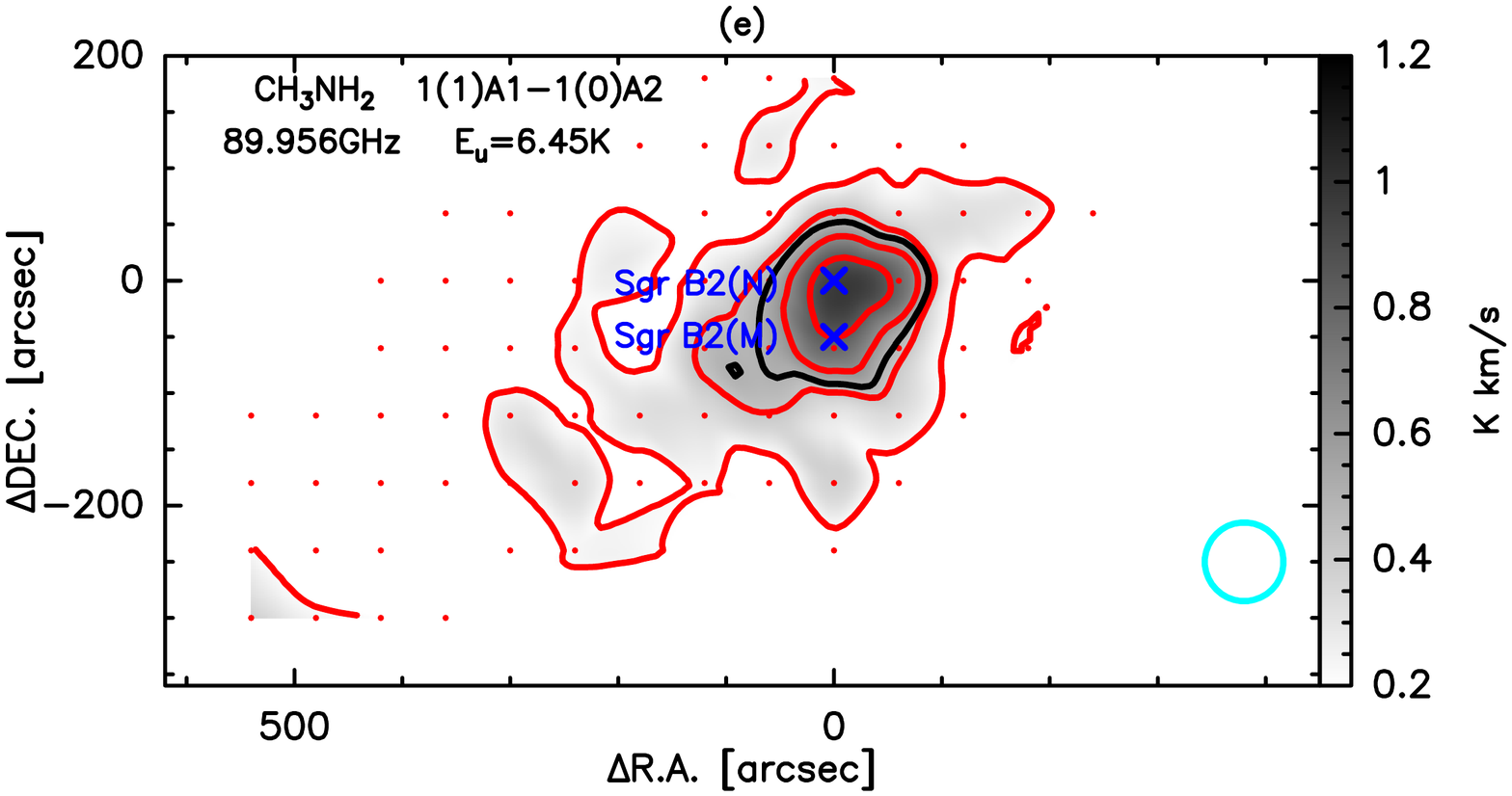}
\caption{(a) Velocity integrated intensity map of ``warm" glycolaldehyde CH$_2$OHCHO 9(1,8)-8(2,7) (contours) overlayed on velocity integrated intensity map of ``cold" glycolaldehyde CH$_2$OHCHO 1(1,0)-1(0,1) observed at 13 GHz\citep{li2017} in grey scale. The contours are from 3$\sigma$ increasing in steps of 3$\sigma$, which corresponds to 0.21 K km s$^{-1}$. The 70 arcsec FWHM ARO beam at 90 GHz is shown by cyan circle, while the 77 arcsec FWHM TMRT beam at 13 GHz is shown by black circle. (b) Velocity integrated intensity map of CH$_3$OCHO 7(2,5)-6(2,4)E (contours) overlayed on velocity integrated intensity map of ``cold" glycolaldehyde CH$_2$OHCHO 1(1,0)-1(0,1) observed at 13 GHz\citep{li2017} in grey scale. The contours are from 5$\sigma$ increasing in steps of 5$\sigma$, which corresponds to 0.35 K km s$^{-1}$. The 70 arcsec FWHM ARO beam at 90 GHz is shown by cyan circle, while the 77 arcsec FWHM TMRT beam at 13 GHz is shown by black circle. (c) Velocity integrated intensity map of t-HCOOH 4(0,4)-3(0,3). The contours are from 5$\sigma$ increasing in steps of 5$\sigma$, which corresponds to 0.35 K km s$^{-1}$. (d) Velocity integrated intensity map of C$_2$H$_5$OH 4(1,4)-3(0,3). The red contours started from and with the step of 1 K km s$^{-1}$. (e) Velocity integrated intensity map of CH$_3$NH$_2$ 1(1,0)-1(0,1). The contours are from 3$\sigma$ increasing in steps of 3$\sigma$, which corresponds to 0.21 K km s$^{-1}$. The black contours represent 50\% of the map peaks, while dots stand for the sampling points in each sub-plot. The 70 arcsec FWHM ARO beam at 90 GHz is shown by cyan circle in each sub-plot.}
\label{fig 2}
\end{figure*}

\begin{figure*}
\centering
\includegraphics[width=2.3in, angle=-90]{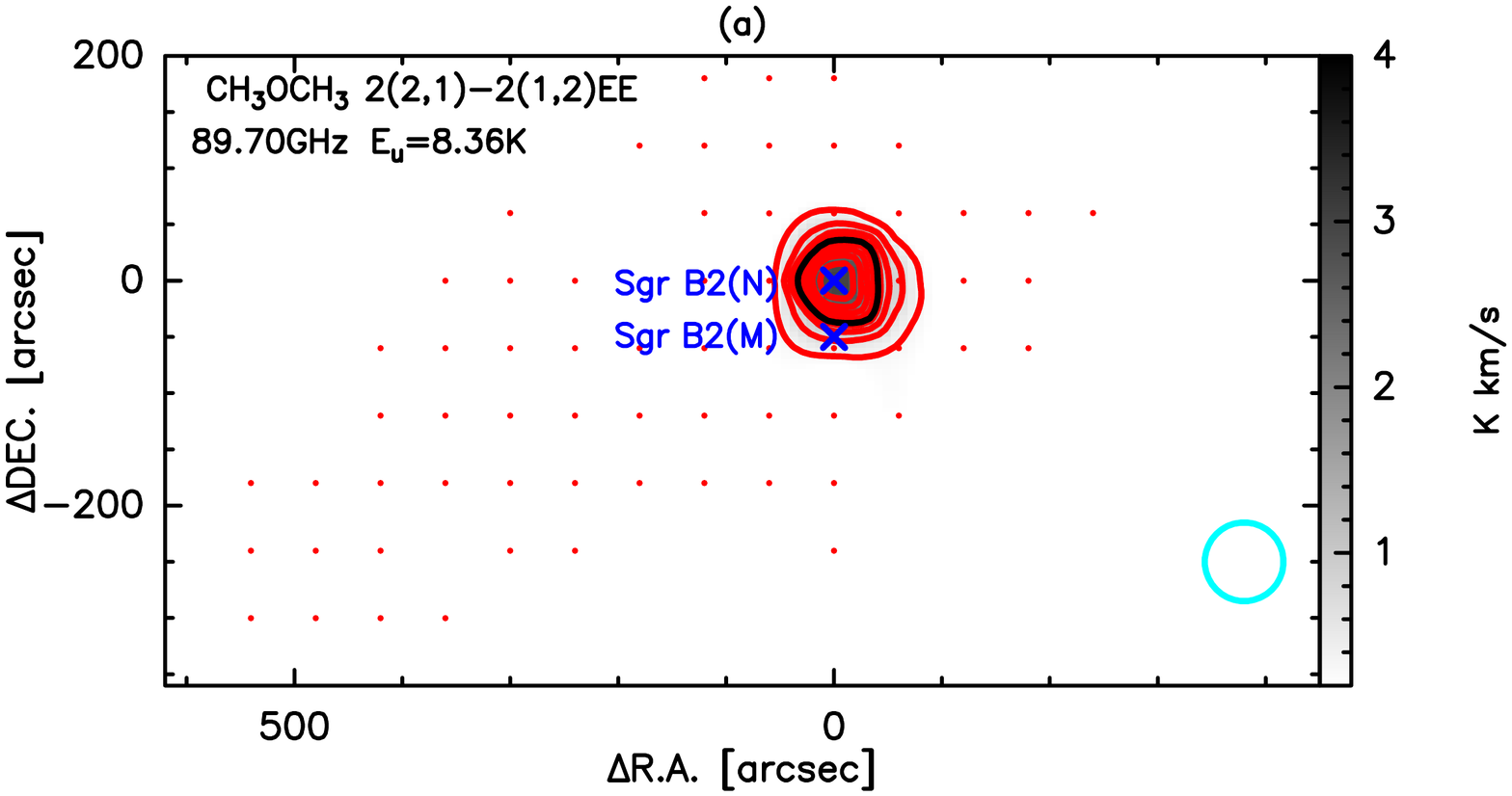}
\includegraphics[width=2.3in, angle=-90]{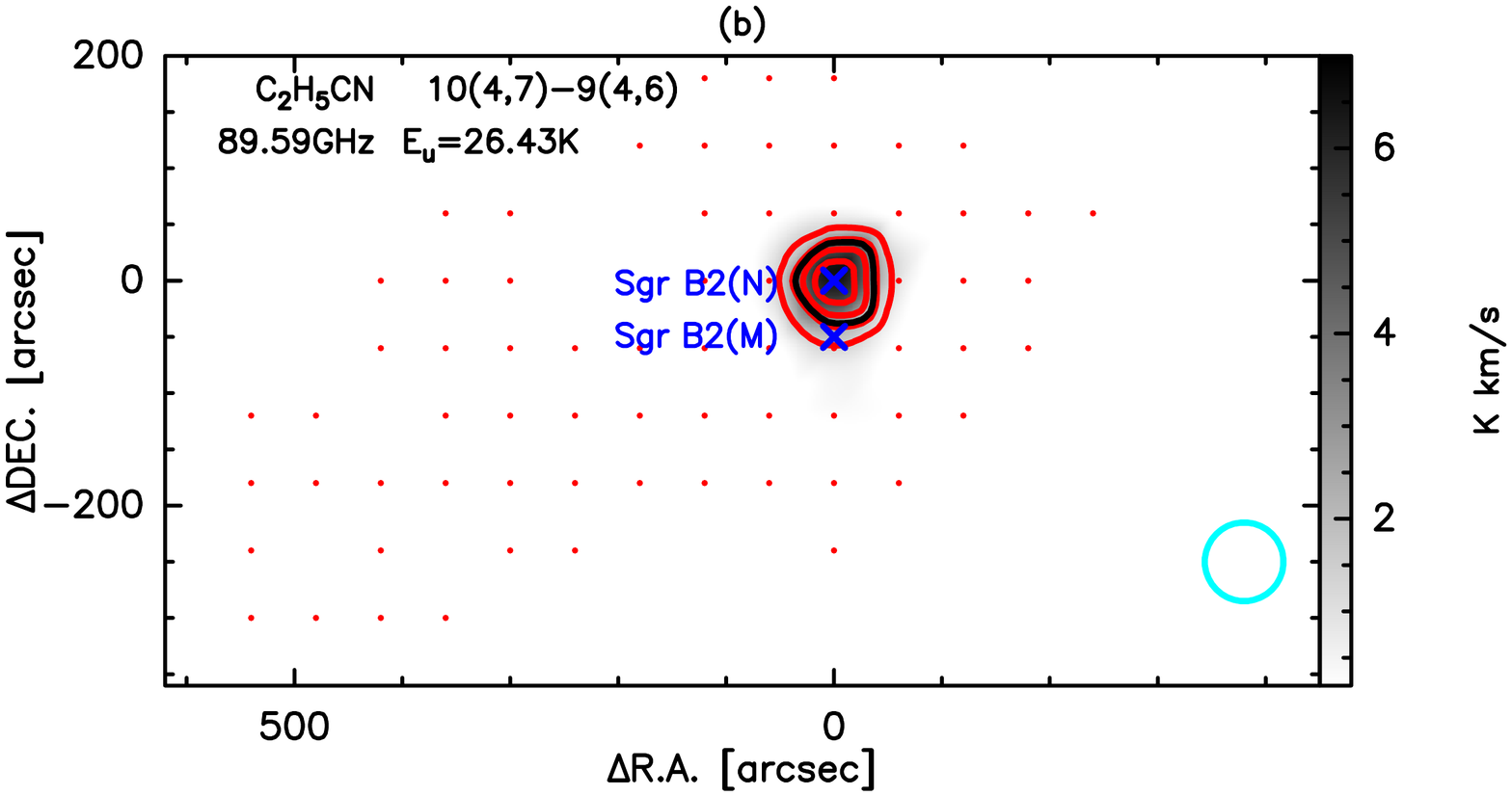}
\includegraphics[width=2.3in, angle=-90]{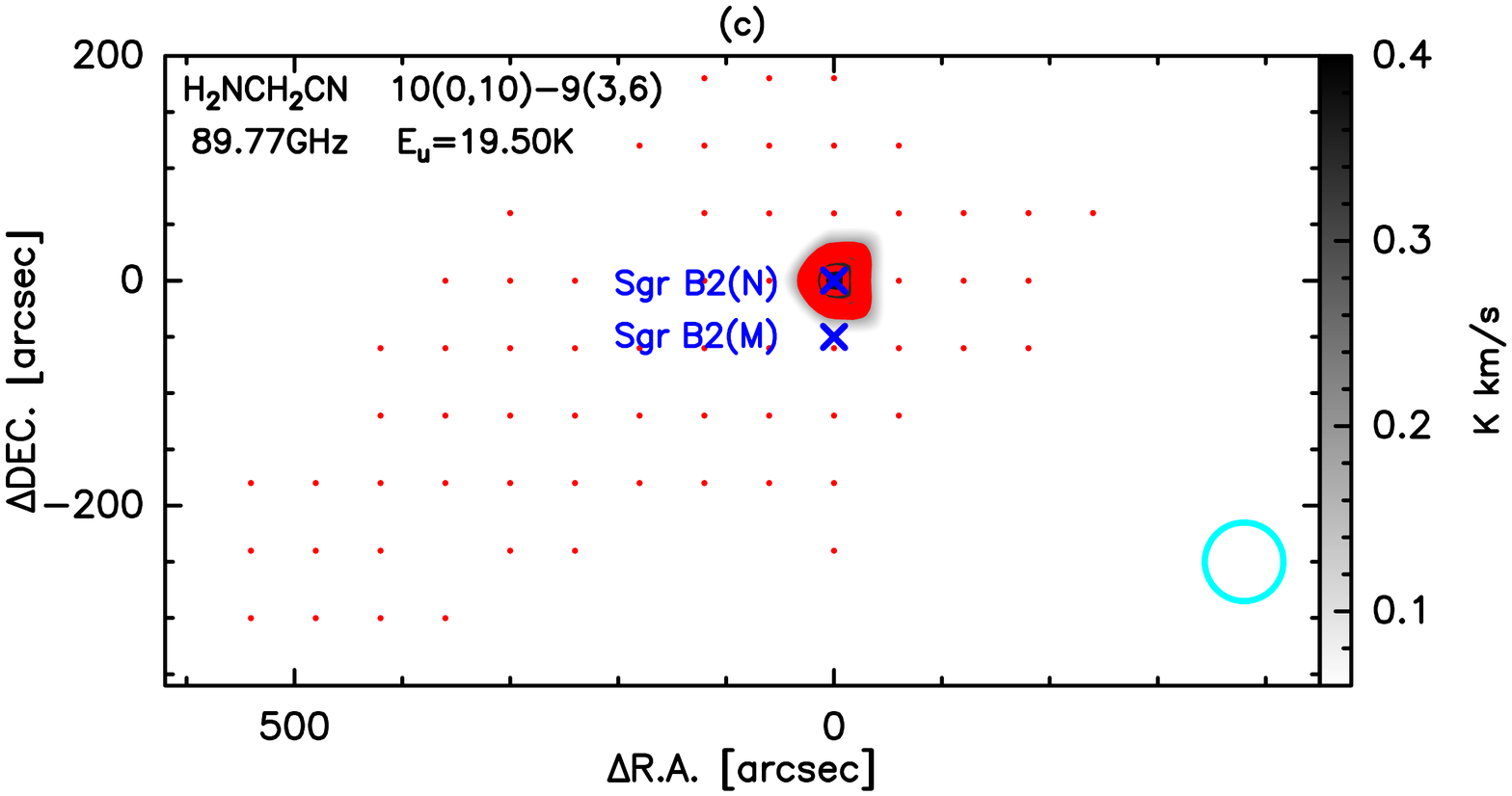}
\caption{(a) Velocity integrated intensity map of CH$_3$OCH$_3$ 2(2,1)-2(1,2)EE observed with ARO 12m telescope. The contours are from 10$\sigma$ increasing in steps of 10$\sigma$, which corresponds to 0.7 K km s$^{-1}$. (b) Velocity integrated intensity map of C$_2$H$_5$CN 10(4,7)-9(4,6) observed with ARO 12m telescope. The contours are from 20$\sigma$ increasing in steps of 20$\sigma$, which corresponds to 1.4 K km s$^{-1}$. (c) Velocity integrated intensity map of H$_2$NCH$_2$CN 10(0,10)-9(3,6) observed with ARO 12m telescope. The contours are from 5$\sigma$ increasing in steps of $\sigma$, which corresponds to 0.04 K km s$^{-1}$. The black contours represent 50\% of the map peaks. Dots stand for the sampling points. The 70 arcsec FWHM ARO beam at 90 GHz is shown by cyan circle in each sub-plot.}
\label{fig 3}
\end{figure*}

The many hazards of searching for new molecules have been pointed out by \cite{snyder2005}. These authors proposed some of the criteria needed for secure molecular identification,
including accurate rest frequencies and consistency among the acquired data set. \cite{halfen2006} extended Criteria in \cite{snyder2005} and applied them to the ARO 12m observations. They suggest that for an accurate identification, there must be strong evidence for emission at all favorable, physically connected transitions over a sufficiently large wavelength range. There cannot be `missing' favorable lines. Based on these criteria, CH$_2$OHCHO 9(1,8)-8(2,7) (89.868639 GHz), the glycolaldehyde transition used in this work was found to be strong and relatively clean. \cite{xue2019} proposed a quantitative method to identify weak and mostly uncontaminated transitions. They used P factor, which is the product of the observed and simulated line profiles, to characterizing the frequency agreement. They used D factor, which is the ratio between the difference of the integrated intensities of the compared spectra and the maximum of the two subtracted from 1, to measure the difference of the line intensities. For CH$_2$OHCHO 9(1,8)-8(2,7) at 89.868639 GHz, the P factor is 94\%, while the D factor is 16.8\%. For CH$_3$OCHO 7(2,5)-6(2,4) E at 90.145 GHz, the P factor is 99.8\%, while the D factor is 25.4\% (see appendix in \cite{xue2019}). Based on the P factors, both of these two transitions are assigned to be unblended transition candidates. The low D factors are likely to be caused by the high excitation temperature (190 K) adopted by \cite{xue2019}. Transitions of other molecules have been identified by \cite{belloche2013}. Velocity integrated images of strong and relatively clean transitions, including CH$_2$OHCHO 9(1,8)-8(2,7), CH$_3$OCHO 7(2,5)-6(2,4)E, t-HCOOH 4(0,4)-3(0,3), C$_2$H$_5$OH 4(1,4)-3(0,3) and CH$_3$NH$_2$ 1(1,0)-1(0,1), CH$_3$OCH$_3$ 2(2,1)-2(1,2)EE, C$_2$H$_5$CN 10(4,7)-9(4,6) and H$_2$NCH$_2$CN 10(0,10)-9(3,6) were presented in Figure \ref{fig 2} and Figure \ref{fig 3}. Positions of Sgr B2(N) and Sgr B2(M) are indicated with crosses. The peak integrated intensity on the T$_{mb}$ scale are given in the caption. We could see from Figure \ref{fig 2} and Figure \ref{fig 3} that some molecular line emission concentrate on Sgr B2(N) and Sgr B2(M), and some molecular line emission extend to the eastern part of Sgr B2. 
%Due to the lack of excitation information, it is unclear whether it is caused by chemistry or excitation. 
Based on their spatial distribution, molecules can be classified as either ``extended" that detected not only in Sgr B2(N) and Sgr B2(M), or ``compact" that only detected toward or near to Sgr B2(N) and Sgr B2(M).

The excitation analysis of glycolaldehyde emissions at millimeter and centimeter wavelengths in Sgr B2(N) indicates that a single temperature component was unable to reproduce all the observed spectra \citep{hollis2004}. Thus, two temperature components were invoked: a ``warm" extended glycolaldehyde envelope surrounded by a ``cold" glycolaldehyde halo \citep{hollis2004}. The 89 GHz transitions with a higher upper-level energy (26.43 K) are regarded as coming from the ``warm" phase, while transitions with a lower upper-level energy (1.2 K) are regarded as coming from the ``cold" phase. The ``cold" glycolaldehyde was found to be widespread around the Sgr B2 complex \citep{li2017}. Figure 2a shows a contour map of the velocity-integrated intensities of the ``warm" glycolaldehyde emission obtained with the ARO 12m telescope overlaid on the ``cold" glycolaldehyde emission acquired with the Shanghai Tianma 65m telescope in grey scale \citep{li2017}. The half-power beamwidths were $\sim$70 arcsec and $\sim$77 arcsec for ARO 12 m observations and 65m observations, respectively. The ``cold" glycolaldehyde is distributed across 800 arcsec ($\sim$33 pc in diameter), which is substantially more extended than the ``warm" glycolaldehyde. The ``warm" glycolaldehyde emission shows several peaks, including the north cloud, the peak near Sgr B2(M), and the ridge west of Sgr B2(N). A weak ``warm" glycolaldehyde emission at 89 GHz was also detected toward the eastern component of the Sgr B2 complex. 

\begin{figure*}
\centering
\includegraphics[width=2.3in]{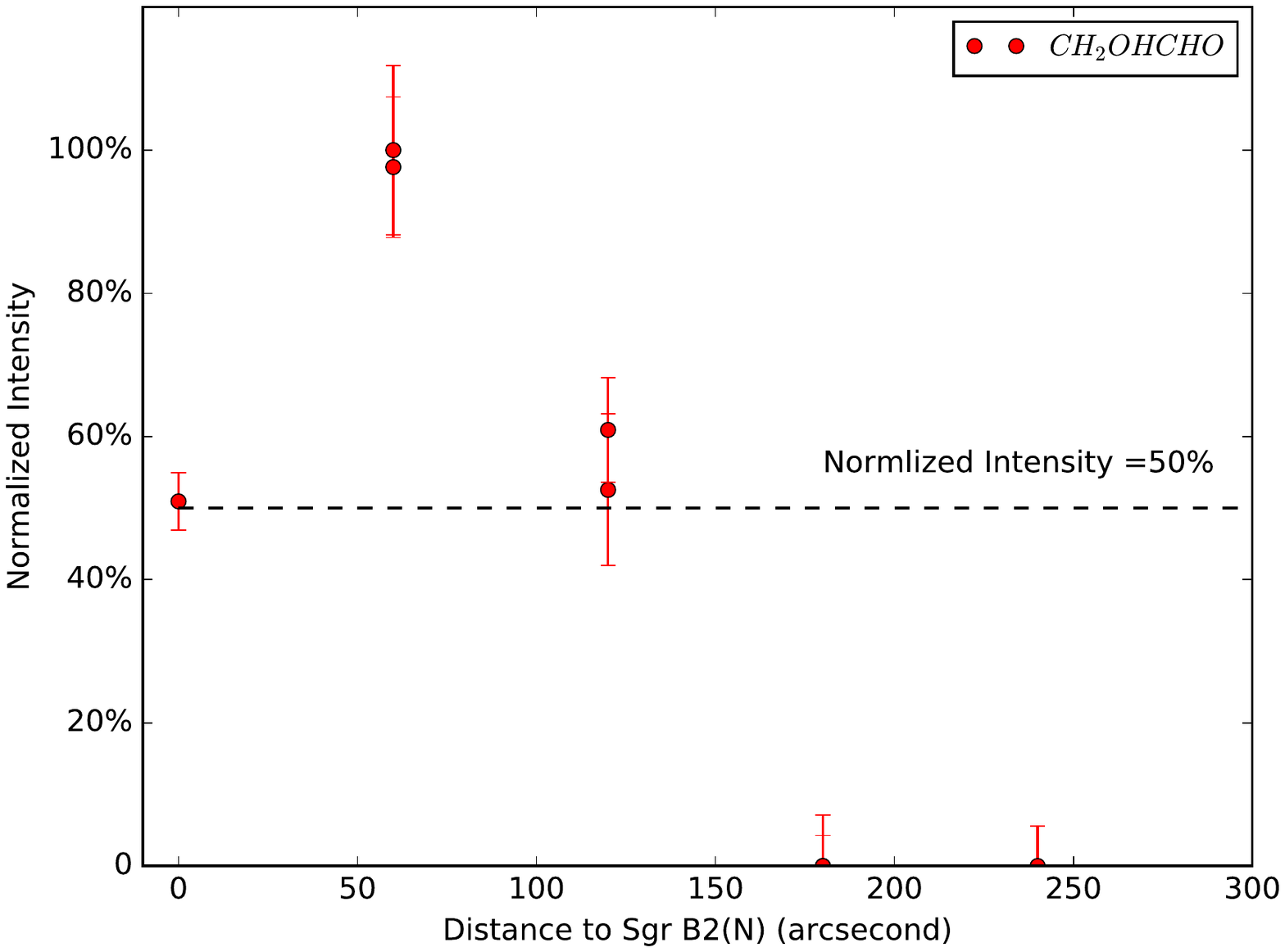}
\includegraphics[width=2.3in]{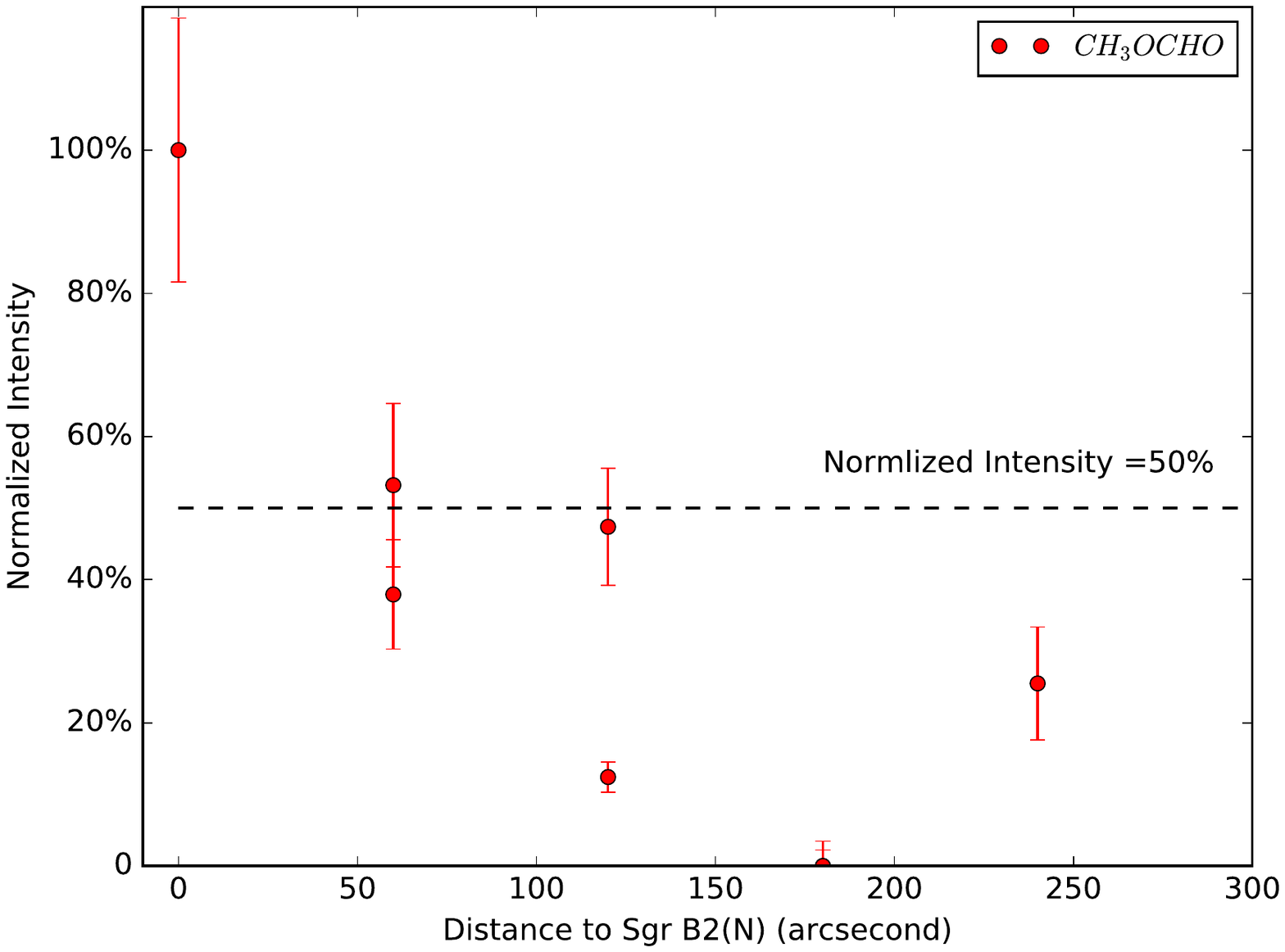}
\includegraphics[width=2.3in]{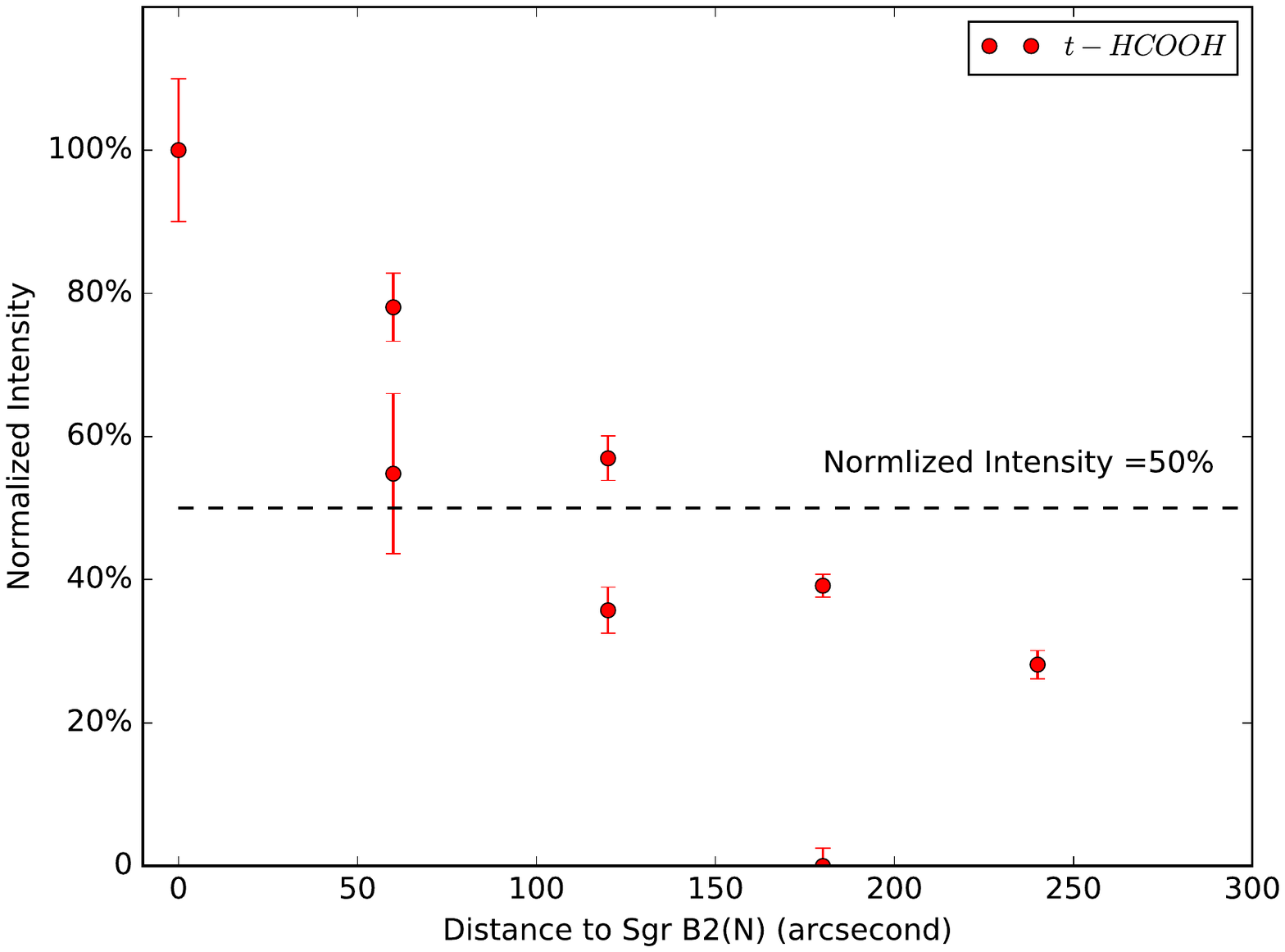}
\includegraphics[width=2.3in]{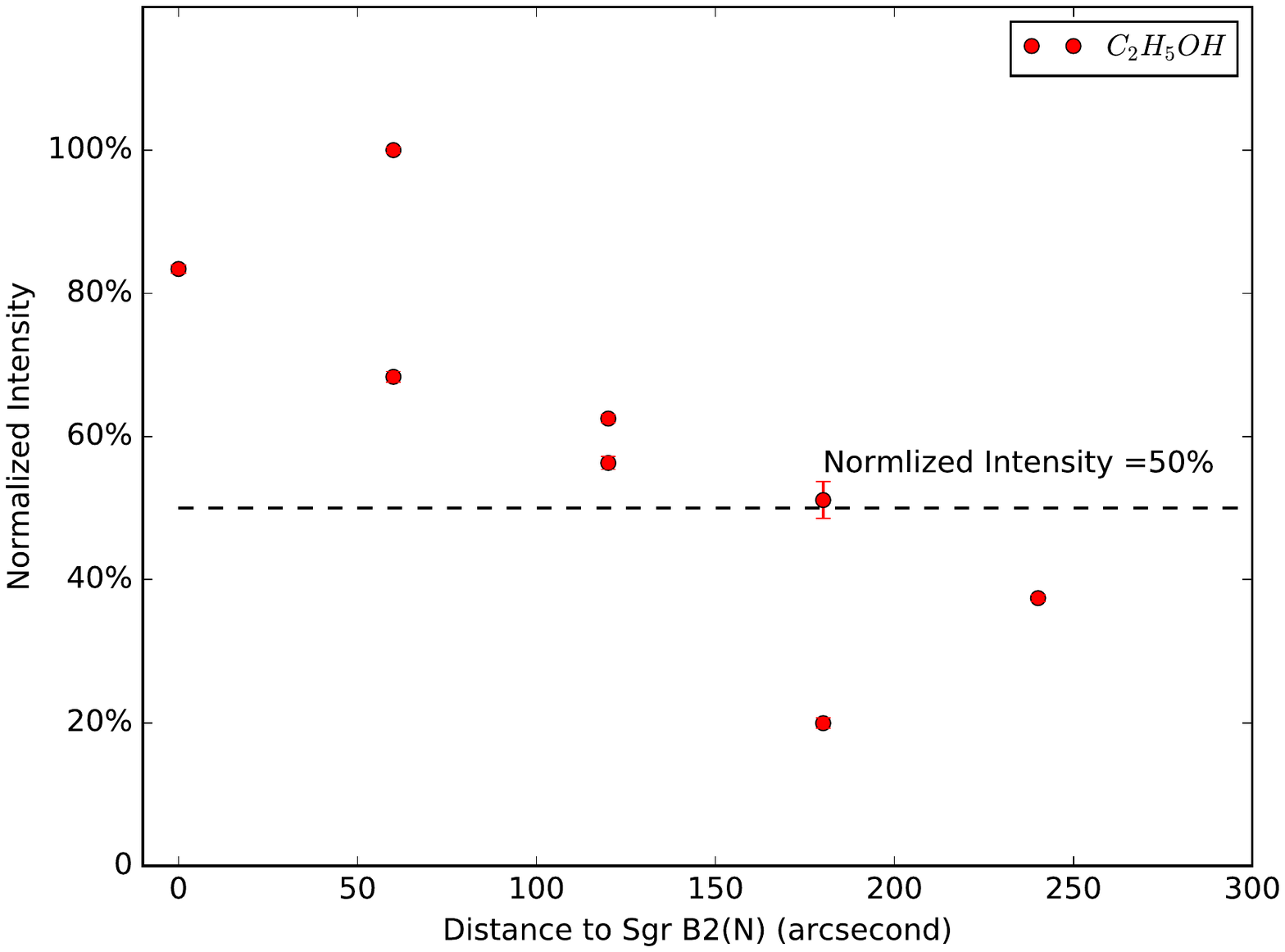}
\includegraphics[width=2.3in]{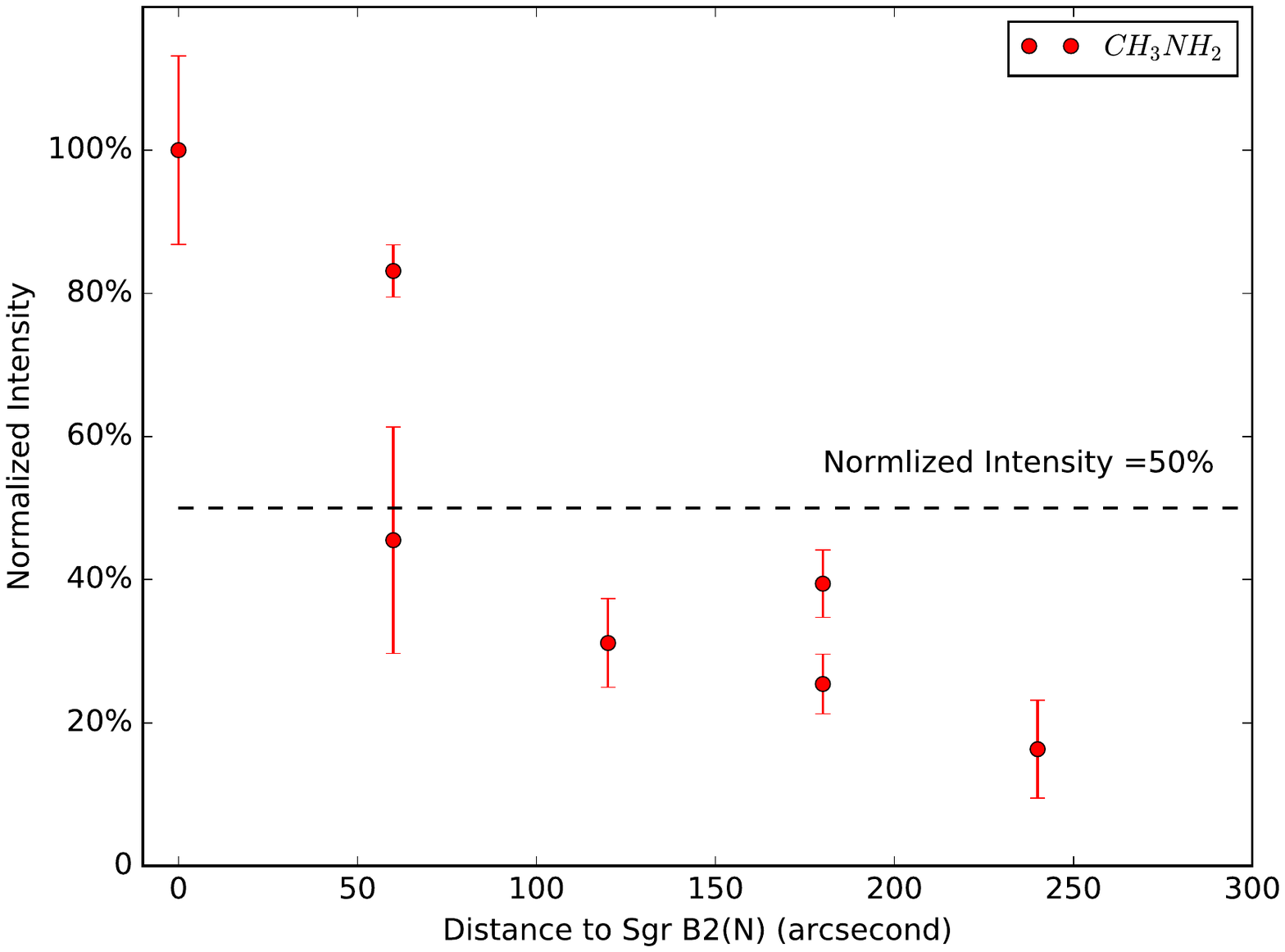}
\includegraphics[width=2.3in]{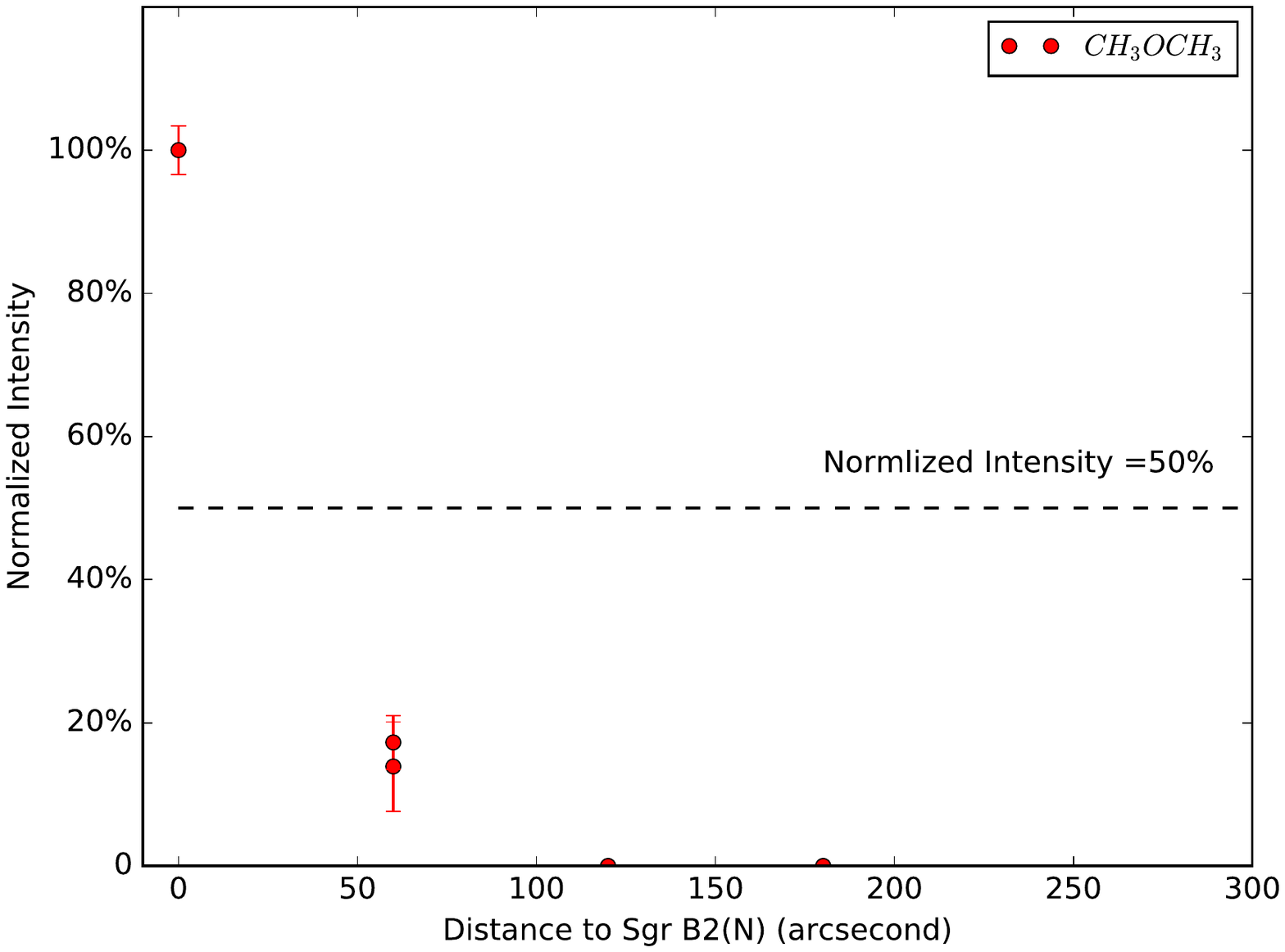}
\includegraphics[width=2.3in]{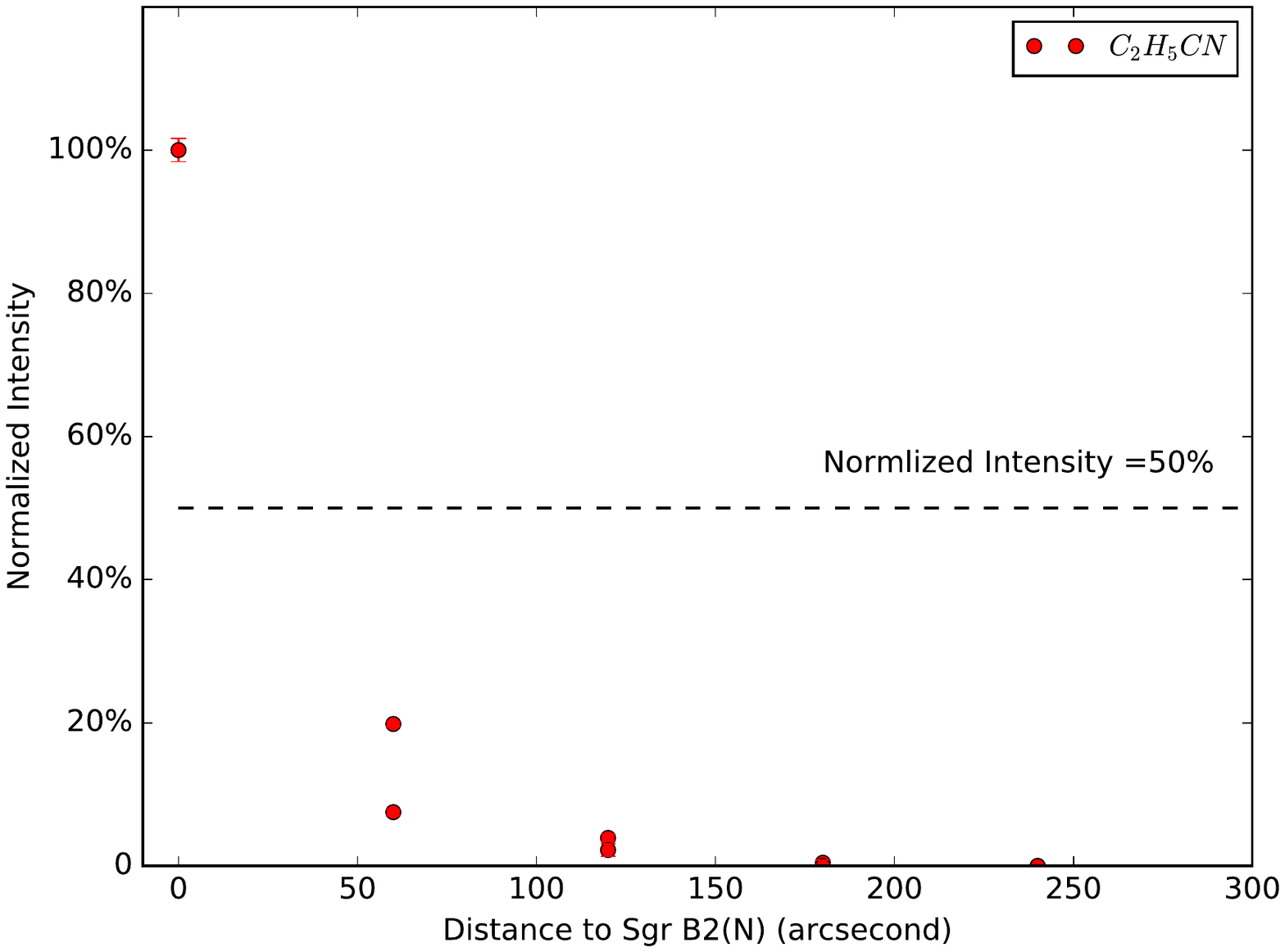}
\vspace{-2 cm}
\caption{Variation of the normalized intensity of observed transitions versus the distance to Sgr B2(N) for sampling points with $\Delta$RA=0. The normalized intensity is obtained through dividing the integrated intensity by the maximum integrated intensity of molecule. We change declination and take a measurement at constant RA ($\Delta$RA=0). 
}
\label{fig 4}
\end{figure*}

According to BIMA observations \citep{hollis2001}, glycolaldehyde is greatly extended in comparison to the ethyl cyanide and dimethyl ether, which are largely confined to the Large Molecule Heimat  source (LMH) \citep{snyder1994, miao1995}, so these molecules were classified as either ``extended" or ``compact" here based on the half-power radius of the spatial distributions in Figures \ref{fig 2} and \ref{fig 3}. 
%Based on the half-power radius of the spatial distributions in Figures \ref{fig 2} and \ref{fig 3}, the molecules can be classified as either ``extended", which are widely spread out, or ``compact", with detections only toward or near to Sgr B2(N) and Sgr B2(M). 
Figure \ref{fig 4} shows variation of the normalized integrated intensity of observed transitions with the distance to Sgr B2(N) for sampling points with $\Delta$RA=0. The normalized intensity is obtained through dividing the integrated intensity by the maximum integrated intensity of molecule. We change the declination and take a measurement at constant RA ($\Delta$RA=0). FWHM, the full width at half maximum, is always used to evaluate the line width of spectral line. Thus we selected 50\% as a metric to evaluate the distribution of molecules. We could see from Figure \ref{fig 4} that the normalized intensity of CH$_3$OCH$_3$, C$_2$H$_5$CN and H$_2$NCH$_2$CN quickly decrease below 50\% for sampling points away from Sgr B2(N). The half-power radius of these molecules are smaller than the sampling interval (1\arcmin). Thus these three molecules are classified as ``compact'' molecules. H$_2$NCH$_2$CN was only detected in Sgr B2(N) in our observations, thus it is also classified as ``compact" molecules. On the other hand, the normalized intensity of CH$_2$OHCHO, CH$_3$OCHO, t-HCOOH, C$_2$H$_5$OH and CH$_3$NH$_2$ vary slowly as the distance to Sgr B2(N), thus these molecules are classified as ``extended" molecules. In order to investigate whether only the strongest lines are extended, the ``extended" line profiles were colored with red, while the ``compact" line profiles were colored with blue in spectra of Sgr B2N (Figure 1 upper panel). As is shown in Figure 1 upper panel, the strongest emissions come from C$_2$H$_5$CN, C$_2$H$_5$OH, t-HCOOH and CH$_3$OCH$_3$. Among them, C$_2$H$_5$CN and CH$_3$OCH$_3$ emission are confined to Sgr B2N and Sgr B2M. Thus it is sure that not all the strongest lines are the most extended. It is noted that CH$_2$OHCHO peaks away from Sgr B2N, which seems to differ from most COMs presented in this paper. The line blending makes it difficult to obtain the intensity of CH$_2$OHCHO accurately. We have mapped CH$_2$OHCHO around Sgr B2 with IRAM 30m telescope with better spatial and spectral resolution to further investigate this issue. The preliminary result is consistent with ARO 12m result present here. Detailed chemical model is needed to explain the spatial distribution of glycolaldehyde in this region.

%Chemical segregation of complex organic O-bearing molecules has also been found in Orion KL \citep{tercero2018}, the closest high-mass star-forming region and another good target for studies of complex organic molecules. 

%The difference between ``extended" and ``compact" molecules is likely due to the fact that the physical conditions of hot cores or nearby regions are different from those of cold regions. When the temperature is relatively high, radicals with large diffusion barriers become mobile and can form more complicated molecules. Higher temperatures within or near the hot cores can also help molecules to overcome the barriers and desorb into the gas phase \citep{garrod2006, garrod2008}.

\subsection{``Extended" Molecules} 

{\it Glycolaldehyde CH$_2$OHCHO}: CH$_2$OHCHO, which is a sugar-related molecule, can react with propenal to form ribose - a central constituent of RNA \citep{sharma2016}. CH$_2$OHCHO, was first detected in Sgr B2(N) based on the emission spectra in the millimeter-wave range observed with the NRAO 12m telescope \citep{hollis2000}. The CH$_2$OHCHO 8(0,8)-7(1,7) emission at 89 GHz was regarded to come from ``warm" glycolaldehyde, while the 13 GHz emission was regarded to come from ``cold" glycolaldehyde \citep{hollis2004}. The distribution of ``warm" glycolaldehyde is extended and seems to be comprised of several emission cores (see Figure \ref{fig 2}). Sensitive interferometric observations are needed to resolve and study these cores in detail. The ``cold" glycolaldehyde has been found to be widespread around the Sgr B2 complex \citep{li2017}. The different spatial distribution of ``warm" and ``cold" glycolaldehyde seems to be caused by different excitation conditions, including temperature and density of molecular gas. 
%Recently, \citep{xue2019} argued that CH$_2$OHCHO 8(0,8)-7(1,7) transition at 89 GHz may be contaminated since the observed intensity of this line is much higher than the simulated values with excitation temperature of 190 K. They proposed that the CH$_2$OHCHO emission should concentrate on the hot cores with excited temperatures above 100 K based on ALMA observations. However, the extended emission of CH$_2$OHCHO has been directly observed by \citep{li2017} at 13 GHz. The adopted excitation temperature in \citep{xue2019} seems to be too high to reproduce the CH$_2$OHCHO 8(0,8)-7(1,7) transition at 89 GHz. The CH$_2$OHCHO 8(0,8)-7(1,7) emission is likely to come from extended envelope with temperatures of about 50 K \citep{hollis2000}, instead of the hot cores.  

{\it Methyl formate CH$_3$OCHO}: ARO observations indicate that the spatial distribution of methyl formate emission significantly differ from both the ``warm" and the ``cold" glycolaldehyde. It seems to peak toward Sgr B2(N) and Sgr B2(M), and decrease toward the surroundings. Weak emission was seen at the eastern part of Sgr B2 complex. 

{\it Formic acid t-HCOOH}: t-HCOOH is a key organic molecule as the carboxyl group (C(=O)OH) is one of the main functional groups of amino acids (the structural units of proteins). This species is involved in a chemical route leading to glycine, the simplest amino acid \citep{redondo2015}. ARO observations indicate that the spatial distribution of t-HCOOH extends to the eastern component and peaks at Sgr B2(N).
 
{\it Ethanol C$_2$H$_5$OH}:  C$_2$H$_5$OH was first detected toward Sgr B2 \citep{zuckerman1975}. This molecular was found to be present only in the dense ($\geq 10^6$ cm$^{-3}$) and hot ($\geq $100 K) cores associated with newly formed massive stars \citep{weaver2017, qin2010, bisschop2007, martin2001}. \cite{martin2001} found that C$_2$H$_5$OH emission in Sgr B2 is widespread. They proposed that C$_2$H$_5$OH formed in grains and released to gas phase by shocks in the last $\sim 10^5$ yr. Results present here provide further evidence for the presence of extended ethanol around Sgr B2. The C$_2$H$_5$OH emission peaks at the north of Sgr B2(N). 

{\it Methylamine CH$_3$NH$_2$}: CH$_3$NH$_2$ was first detected in Sgr B2(OH) \citep{kaifu1974}. It was also detected in the southern region of Sgr B2 \citep{halfen2013}. CH$_3$NH$_2$ is thought to be potential interstellar precursors to the amino acid glycine, NH$_2$CH$_2$COOH. The reaction of CH$_3$NH$_2$ with CO$_2$ in water ice has been shown to yield NH$_2$CH$_2$COOH after UV irradiation \citep{bossa2009, lee2009}. Our observations indicate that CH$_3$NH$_2$ emission peaks at Sgr B2(N), and extends to the eastern part of Sgr B2 complex, which is in agreement with previous results.  

\subsection{``Compact" Molecules} 

{\it Dimethyl ether CH$_3$OCH$_3$}: The CH$_3$OCH$_3$ emission is detected toward Sgr B2(N) and a few points near to Sgr B2(N).  
Though CH$_3$OCH$_3$ is isomer of C$_2$H$_5$OH, the distribution of these two molecules are strikingly different, implying that they have different chemical route, which is consistent with observations in  W51 \citep{rong2015}. As C$_2$H$_5$OH is thought to be produced by grain-surface reactions, CH$_3$OCH$_3$ is likely to form via gas-phase chemistry. The spatial distribution of CH$_3$OCH$_3$ also differ from those of CH$_3$OCHO, which seem to be different from observations in massive star-forming regions in Galactic disk. Striking similarity was found in the spatial distribution, temperature and column densities of these two molecules in star-forming regions in Galactic disk \citep{brouillet2013, jaber2014, rong2015}. CH$_3$OCH$_3$ has been proposed as a precursor molecule to CH$_3$OCHO \citep{balucani2015}.
The obvious difference in spatial distribution of CH$_3$OCH$_3$ and CH$_3$OCHO observed in Sgr B2 suggests that the formation mechanism of these two molecules might differ in the Galactic center and Galactic disk sources. 

{\it Ethyl cyanide C$_2$H$_5$CN}: Previous observations of C$_2$H$_5$CN in Sgr B2 has shown that this molecule is confined to a small region, and likely to be located in the hot, dense core of the star-forming region \citep{miao1997}. Our mapping results show that C$_2$H$_5$CN was only detected toward Sgr B2(N), and four points near to Sgr B2(N), which agree with previous studies. 

{\it Amino acetonitrile H$_2$NCH$_2$CN}: H$_2$NCH$_2$CN was first detected in Sgr B2(N) \citep{belloche2008}. The source size was estimated to be about 2'' FWHM. No evidence for colder, more extended emission was found. Amino acetonitrile may well be a direct precursor of glycine. In our observations, H$_2$NCH$_2$CN was only detected toward Sgr B2(N).

\subsection{Column densities}

According to IRAM 30m observations of COMS \citep{requena2006}, $T_{rot}$ derived from C$_2$H$_5$OH is 72.6 K for Sgr B2N, and 55.4 K for Sgr B2M. $T_{rot}$ derived from C$_2$H$_5$OH range from 9-14 K for other Galactic center clouds, $T_{rot}$ derived from CH$_3$OCHO range from 8-16 K for other Galactic center clouds. Assuming the excitation temperature of 50 K for Sgr B2N and Sgr B2(M), and 14 K for other positions, we calculate the total column densities of ``warm" CH$_2$OHCHO, CH$_3$OCHO, C$_2$H$_5$OH and t-HCOOH for the i, j-th grid in the position-position space $N_{ij}$ with the expression reported by \cite{hollis2004}:
\begin{eqnarray}
N_{ij}=\frac{3kQe^{E_u/kT_S}}{8 \pi^3 \nu S \mu^2 } \times \frac {\frac{1}{2} \sqrt{\frac{\pi}{ln 2}} \frac{\triangle T_A^* \triangle V}{\eta_B}} {1-\frac{e^{h\nu/kT_S}-1}{e^{h\nu/KT_{bg}}-1}} ,
\end{eqnarray}
in which $k$ is the Boltzmann constant in erg K$^{-1}$, $\frac{\triangle T_A^* \triangle V}{\eta_B}$ is the observed line integrated intensity in K Km s$^{-1}$, $\nu$ is the frequency of the transition in Hz, and $S\mu^2$ is the product of the total torsion-rotational line strength and the square of the electric dipole moment. $T_{S}$ and $T_{bg}$ (=2.73 K) are the excitation temperature and background brightness temperature, respectively. $E_u/k$ is the upper level energy in K. The partition function, $Q$, was estimated by fitting the partition function at different temperatures given in CDMS \citep{muller2005}. Values of $E_u/k$ and $S\mu^2$ are also taken from CDMS.

%{\bf \citep{belloche2013} found that the background temperatures of 5.2 K and 5.9 K at 3 mm atmospheric window for Sgr B2(N) and Sgr B2(M). In order to test the effect of the value of $T_{bg}$, we calculate assume background temperature of 5 K. Then, the percentage differences of N(HC3N) and Texc with respect to the values obtained without optical depth correction are about 19 percent higher and 5 percent lower, respectively. In comparison with the LTE analysis of HC3N, the results obtained assuming non-LTE conditions are relatively in good agreement with those obtained by the excitation diagram for the v=0 state (see Sect. 4.2.1). The percentage differences of column density and temperature with respect to the values obtained in LTE conditions for the v=0 lines with optical depth correction are about 6 percent higher and 53 percent lower, respectively, while the difference with re- spect to the values without optical depth correction is only about 1 percent higher and 28 percent lower, respectively.}

Because of effects of beam dilution, we did not calculate the column densities of ``compact" molecules like C$_2$H$_5$CN and CH$_3$OCH$_3$. In the below discussion, we focus only on oxygen-bearing molecules showing extended emission, including ``warm" CH$_2$OHCHO, CH$_3$OCHO, C$_2$H$_5$OH and t-HCOOH. 

We made use of the H$_2$ column density obtained with BGPS 1.1 mm data \citep{bally2010} for abundance estimation of molecules. For each position, we searched for nearest clump identified with BGPS data, and obtained the corresponding hydrogen column density. The H$_2$ column densities range from $1.1\times10^{23}$ cm$^{-2}$ to $1.0\times10^{25}$ cm$^{-2}$ in our observing region, and peak toward Sgr B2(N) and Sgr B2(M).

\section{Discussions}

\begin{figure*}
\centering
\includegraphics[width=3.0in]{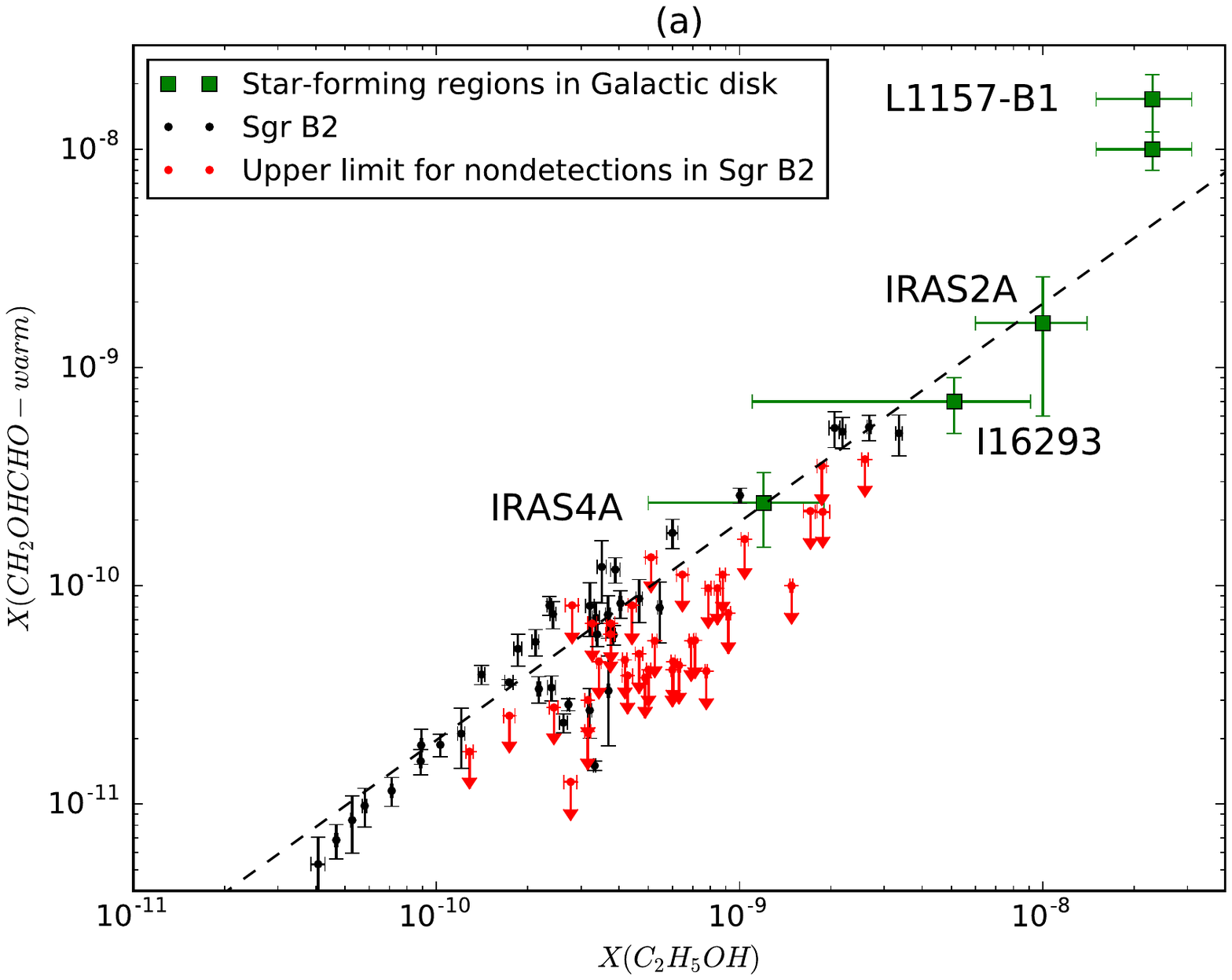}
\includegraphics[width=3.0in]{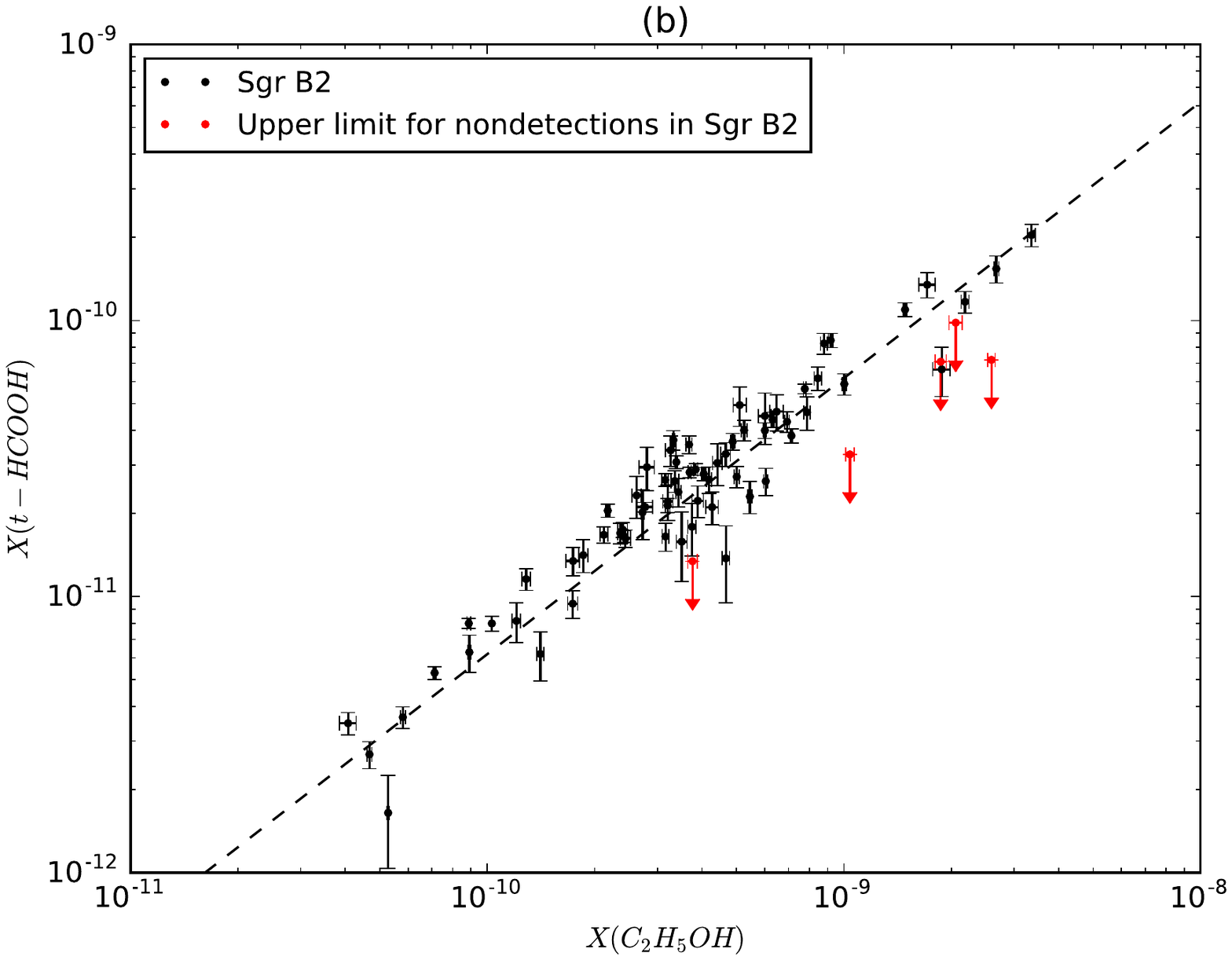}
\includegraphics[width=3.0in]{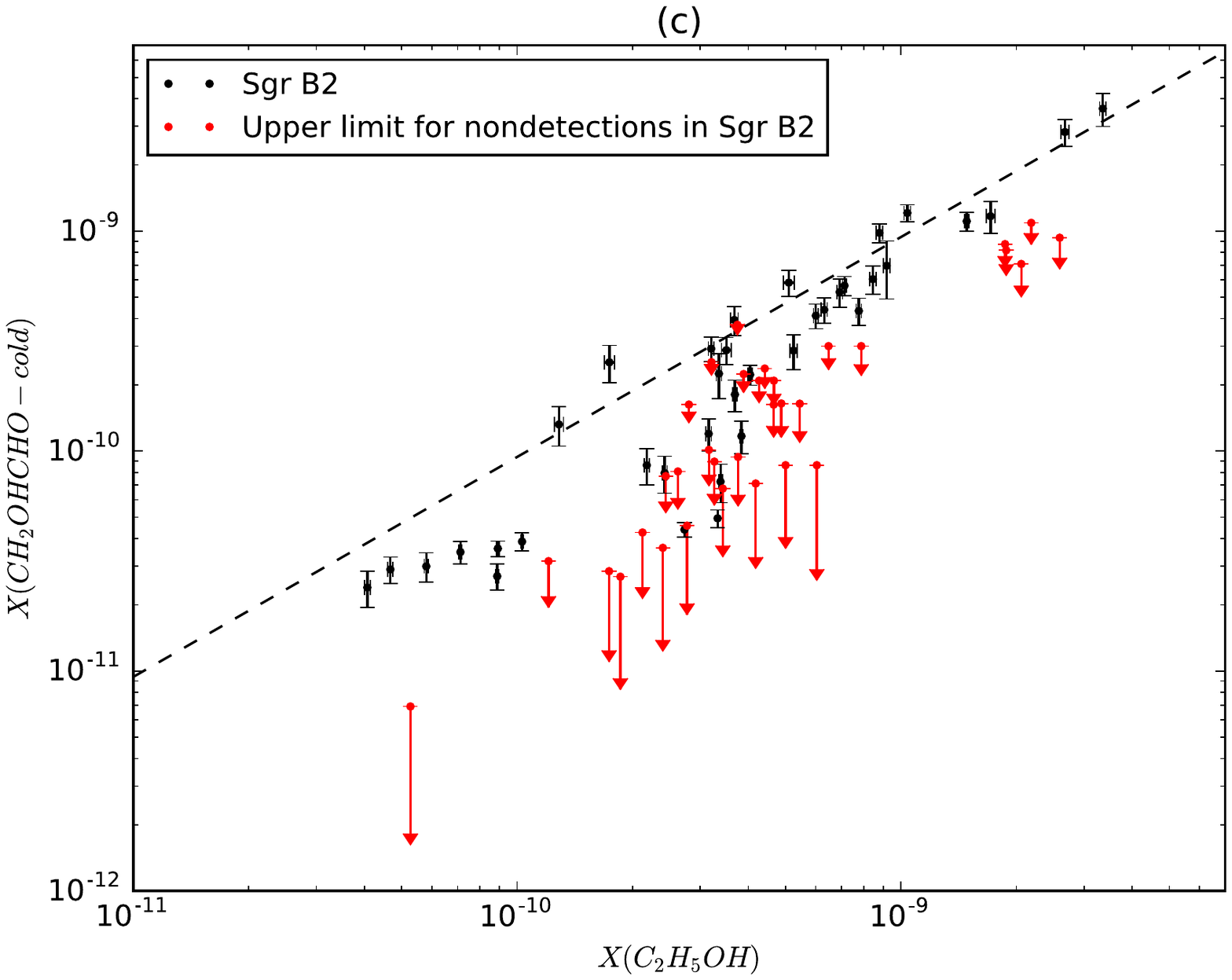}
\caption{(a) ``Warm" glycolaldehyde abundance as a function of ethanol abundance for different positions in Sgr B2. The relation $X(CH_2OHCHO-warm)=0.2X(C_2H_5OH)$ is marked by the black dashed line. (b) Formic acid abundance as a function of ethanol abundance for different positions in Sgr B2. The relation $X_{t-HCOOH}=0.06X_{C_2H_5OH}$ is marked by the black dashed line. (c) ``Cold" glycolaldehyde abundance as a function of ethanol abundance for different positions in Sgr B2. The relation $X_{CH_2OHCHO-cold}=0.94X_{C_2H_5OH}$ is marked by the black dashed line.
}
\label{fig 5}
\end{figure*}

 A comparison between the abundances of ethanol and ``warm" glycolaldehyde relative to that of $N(H_2)$ for each position is shown in Figure \ref{fig 5}. 
 %Figure 5 (right) shows integrated intensity map of C$_2$H$_5$OH shown in contours overlaid on integrated intensity of `cold' glycolaldehyde shown in grey scale. 
We found that $X(CH_2OHCHO-warm)=0.2X(C_2H_5OH)$, with a correlation coefficient of 0.96. This result agrees well with the relation found for star-forming regions in the Galactic disk: L1157-b1 \citep{lefloch2017}, IRAS 16293-2422 \citep{jorgensen2016}, IRAS 2A and IRAS 4A \citep{taquet2015} (see Figure \ref{fig 3}). The strong correlation between ethanol and ``warm" glycolaldehyde suggests that the ``warm" glycolaldehyde may be chemically related to ethanol. Recently, theoretical studies have shown that a new gas-phase scheme of reactions, involving ethanol as a parent molecule, can lead to the formation of glycolaldehyde and formic acid \citep{skouteris2018}. This model can well explain the abundance correlation between ``warm" glycolaldehyde and ethanol. This model is further supported by the strong correlation between the abundances of ethanol and formic acid relative to that of $N(H_2)$ (Figure \ref{fig 5}), with a correlation coefficient of 0.92. The correlation between the abundances relative to H$_2$ of ethanol, ``warm" glycolaldehyde and formic acid suggests that ``warm" glycolaldehyde and formic acid are chemically related to ethanol and are possibly produced by ethanol via gas-phase reactions \citep{skouteris2018}. However, we couldnot exclude the possibility that these molecules are co-spatial in the same gas and trace a warm gas phase. Interferometric observations of a large sample of interstellar sources could help to investigate whether ethanol, formic acid and ``warm" glycolaldehyde are co-spatial and constrain their formation pathways \citep{xue2019}. 
 
Figure \ref{fig 5} shows a comparison between the abundances relative to H$_2$ of ethanol and ``cold" glycolaldehyde for each position. 
%Figure 2 (right) shows integrated intensity map of C$_2$H$_5$OH shown in contours overlaid on integrated intensity of `cold' glycolaldehyde shown in grey scale. 
The abundance of ``cold" glycolaldehyde was observed at the 13 GHz \citep{li2017} range from $10^{-11}$ to $10^{-9}$, which is similar to the value for ethanol. We found that $X_{CH_2OHCHO-cold}=0.94X_{C_2H_5OH}$, with a correlation coefficient of 0.98. Theoretical studies show that gas-phase chemistry cannot produce the observed abundance of ethanol, even under the most favorable conditions \citep{charnley1995}. Infrared observations also indicate that interstellar ethanol can be formed by grain-surface reactions significantly more efficiently than by gas-phase chemistry \citep{sandford1991}; thus, ethanol is thought to be a good tracer of dust chemistry. The good correlation ($\sim$1) between the abundances of ethanol and ``cold" glycolaldehyde indicate that the ``cold" glycolaldehyde should also be produced by grain-surface reactions. Similar to ethanol, the desorption mechanism for the ``cold" glycolaldehyde should be the large-scale shocks in the Galactic center region \citep{martin2001}. 

Laboratory studies show that glycolaldehyde always forms on icy grains at low temperatures, regardless of UV-photon absorption (i.e. energetic processes), and methyl formate cannot form efficiently in the absence of energetic processes \citep{chuang2017}. Overabundant methyl formate would mean that additional UV photons to low-temperature surface chemistry. In contrast, under-abundant methyl formate would mean that only the low-temperature process is important in the production of complex organic molecules. This view agrees well with the results presented here, in which methyl formate emission is concentrated on Sgr B2(N) and Sgr B2(M), where massive star-forming activity takes place and UV-photons are abundant. In contrast, ``cold" glycolaldehyde emission is very widespread and extends to the eastern part of the Sgr B2 complex, where no star-forming activity has been found and the UV radiation should be substantially weaker. Thus, the methyl formate and ``cold" glycolaldehyde near to Sgr B2(N) and Sgr B2(M) likely form on icy grains under UV radiation, while ``cold" glycolaldehyde in regions away from the HII region likely form on icy grains at low temperature. 

\section{SUMMARY}

We have carried out large-scale mapping observations of a series of COMs toward Sgr B2 with the ARO 12m telescope. We obtained large-scale spatial distribution of COMs, including CH$_2$OHCHO, CH$_3$OCHO, t-HCOOH, C$_2$H$_5$OH and CH$_3$NH$_2$, CH$_3$OCH$_3$, C$_2$H$_5$CN, in Sgr B2. The main results of this work include:

1.  The spatial distribution of complex organic molecules can be classified as either ``extended" or ``compact". The ``extended" molecules include CH$_2$OHCHO, CH$_3$OCHO, t-HCOOH, C$_2$H$_5$OH and CH$_3$NH$_2$, while the ``compact" molecules include CH$_3$OCH$_3$, C$_2$H$_5$CN, and H$_2$NCH$_2$CN. 

2. The spatial distribution of CH$_3$OCHO obviously differ from CH$_3$OCH$_3$, which have been observed to have similar spatial distribution and column densities to that in star-forming regions of Galactic disk. These results suggest that the formation mechanisms of these two molecules in Galactic center may differ from that in star-forming regions of Galactic disk. 

3. These ``compact" molecules likely to be produced under strong UV radiation, while ``extended" molecules likely to be formed under low-temperature, via gas-phase or grain surface reactions.

4. We found evidence for an overabundance of CH$_2$OHCHO compared to that expected from the gas-phase model, which indicates that grain-surface reactions are necessary to explain the origin of CH$_2$OHCHO in Sagittarius B2. 

Our measurements demonstrate the necessity of grain-surface chemistry for the production of COMs, such as glycolaldehyde in Sgr B2. The results open up an exciting opportunity to study COMs in the circumnuclear disk of starburst galaxies with interferometers including ALMA, the Next Generation Very Large Array (ngVLA) and SKA. Future high-sensitivity interferometric observations possess the potential to probe COM emissions to investigate whether CMZ clouds could serve as a template for the nuclei of starburst galaxies in the nearby and distant universe \citep{kauffmann2017}.

%\begin{figure*}
%	\centering
%\includegraphics[width=3.2in]{NGC6240_CN.eps}
%\includegraphics[width=3.2in]{VIIZW31_CN.eps}

%\vspace*{-0.2 cm} \caption{CN (1-0) in NGC 6240 and VII ZW 31. The  rms  is  0.88 mK at the velocity resolution of 18.57 km\,s$^{-1}$ in NGC 6240 and 0.99 mK at the velocity resolution of 18.57 km\,s$^{-1}$. The same nine hyperfine lines  are marked from 1-9 in these two sources: 1, CN 1-0 J=3/2-1/2 F=1/2-3/2 at 113.520 GHz; 2, CN 1-0 J=3/2-1/2 F=3/2-3/2 at 113.509 GHz; 3, CN 1-0 J=3/2-1/2 F=1/2-1/2 at 113.4996 GHz; 4, CN 1-0 J=3/2-1/2 F=5/2-3/2 at 113.491 GHz; 5, CN 1-0 J=3/2-1/2 F=3/2-1/2 at 113.488 GHz;   6, CN 1-0 J=1/2-1/2 F=3/2-3/2 at 113.191 GHz; 7, CN 1-0 J=1/2-1/2 F=3/2-1/2 at 113.171GHz; 8, CN 1-0 J=1/2-1/2 F=1/2-3/2 at 113.144 GHz; 9, CN 1-0 J=1/2-1/2 F=1/2-1/2 at 113.123 GHz. The transition of J=3/2-1/2 F=1/2-3/2 in NGC 6240 and J=3/2-1/2 F=5/2-3/2 in VII Zw 31 is used as reference for the velocities, respectively.
%	\label{fig:NGC6240_CN,VIIZW31_CN}}
  %  \vskip-10pt
%\end{figure*}

\section*{Acknowledgements}

The Kitt Peak 12 Meter is operated by the Arizona Radio Observatory (ARO), Steward Observatory, University of Arizona. We thank ARO staff for assisting with the observations. This work is partially supported by the National Key R\&D Program of China (No. 2017YFA0402604), the Natural Science Foundation of China (11590780, 11590782, 11590784, U1431125, 11773054 and 11903038), the Special Funding for Advanced Users, budgeted and administrated by Center for Astronomical Mega-Science, Chinese Academy of Sciences (CAMS-CAS) and CAS ``Light of West China'' Program.

{}

% Don't change these lines
\bsp	% typesetting comment
\label{lastpage}
\end{document}